# Van der Waals Layered Ferroelectric $CuInP_2S_6$：Physical Properties and Device Applications

Shuang Zhou[1†], Lu You[2†*], Hailin Zhou[2], Yong Pu[1], Zhigang Gui[3], Junling Wang[3*]

1  New Energy Technology Engineering Laboratory of Jiangsu Province & School of Science, Nanjing University of Posts and Telecommunications, Nanjing 210023, China
2  Jiangsu Key Laboratory of Thin Films, School of Physical Science and Technology, Soochow University, Suzhou 215006, China
3  Department of Physics, Southern University of Science & Technology, Shenzhen 518055, China

†These authors contributed equally to this work.
\* Corresponding author. E-mail: jwang@sustech.edu.cn; lyou@suda.edu.cn

**Abstract:** Copper indium thiophosphate, $CuInP_2S_6$, has attracted much attention in recent years due to its van der Waals layered structure and robust ferroelectricity at room temperature. In this review, we aim to give an overview of the various properties of $CuInP_2S_6$, covering structural, ferroelectric, dielectric, piezoelectric and transport properties, as well as its potential applications. We also highlight the remaining questions and possible research directions related to this fascinating material and other compounds of the same family.

## Contents

## 1. Introduction

Over the past decade, van der Waals (vdW) materials possessing functional electric, magnetic, thermal, and optical properties have attracted much interest for the development of next-generation multifunctional devices [1-10]. With the rapid expansion of the vdW materials library, the pursuit of long-range ferroic orders, such as (anti)ferroelectricity [11-15], (anti)ferromagnetism [16-18], ferroelasticity [19] and ferrovalley [20,21], in vdW materials is on the rise. Among them, ferroelectricity is widely exploited for potential applications in memories, capacitors, actuators, and sensors [10,15,22]. Below the Curie temperature ($T_C$), spontaneous ordering of electric dipoles produces macroscopic polarization that can be switched by an external electric field. vdW layered ferroelectric materials have become a promising research branch in condensed matter physics [11-15], among which CuInP$_2$S$_6$ (CIPS) is one of the most representative materials because of its room temperature ferroelectricity [13,23]. Furthermore, its out-of-plane polarization direction is more conducive to nonvolatile memories and heterostructure based nanoelectronics/optoelectronics [24,25]. CIPS belongs to the family of transition metal thio/selenophosphates (TPS), a broad class of vdW layered solids in which metal cations are embedded in the lattice framework of thiophosphate (P$_2$S$_6$)$^{4-}$ or selenophosphate (P$_2$Se$_6$)$^{4-}$ anions. They have the general chemical formula of M$^{4+}$[P$_2$X$_6$]$^{4-}$, [M$^{2+}$]$_2$[P$_2$X$_6$]$^{4-}$, and M$^{1+}$M$^{3+}$[P$_2$X$_6$]$^{4-}$, where X = S/Se [26,27]. Pioneering work on the TPS family can be traced back to 1970s through early 2000s owing to their potential in lithium intercalation for battery applications. For a comprehensive review on the family of TPS compounds, the readers are referred to these references [26,28]. Herein, we focus mainly on the physical properties of CIPS itself. The unusual crystal structure and ferroelectric order endow CIPS with intriguing characteristics and rich physics, which attract much interest in the past few years. [13, 23, 29-41].

## 2. Crystal structure and phase transition

### 2.1. vdW layered structure and spatial instability of Cu ion

In 1995, Maisonneuve et al. first resolved the room-temperature crystal structure of CIPS using laboratory single-crystal X-ray diffraction (sc-XRD) [42]. The compound is a layered vdW material with monoclinic symmetry (Space group *Cc*, 9). The crystal structure of CIPS, similar to that of other Cu$^I$M$^{III}$P$_2$S$_6$ compounds (M$^{III}$ =Cr or V) based on ABC close-packed stacking of sulfur atoms, is defined by the sulfur framework in which the metal cations and P-P pairs fill the octahedral voids (**Fig. 1**). The Cu, In, and P-P pairs form triangular patterns within a layer (**Fig. 1b**) [42]. Bulk crystals are composed of vertically stacked and weakly interacting layers linked by vdW interactions. Owing to the site exchange between Cu and P-P pair from one layer to another, a complete unit cell consists of two adjacent layers to fully describe the material's symmetry (**Fig. 1d**) [42]. Recently, the crystal structure of CIPS was re-examined by different research groups using laboratory sc-XRD and synchrotron powder diffraction, as summarized in **Table 1**. Generally, the derived crystal symmetry and lattice parameters are consistent with that reported in the 1995 paper [42]. The only major difference is the monoclinic $\beta$ angle, which is simply due to the different way of unit-cell selection, as depicted by the dotted lines in **Fig. 1a**.

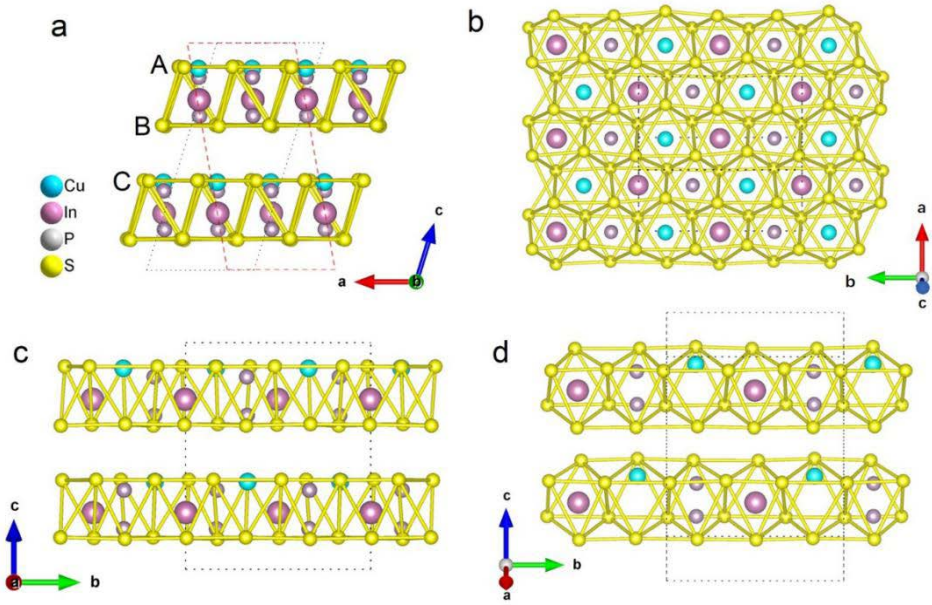

**Fig.1 a-c,** Crystal structure of CIPS viewed from (a) b axis, (b) layer normal ($c^*$ axis) and (c) a axis, respectively. **d**, Perspective view of two [SCu$_{1/3}$In$_{1/3}$(P$_2$)$_{1/3}$S] layers. [42]

| Sample condition | Single crystal 295 K [42] | Single crystal 296 K [41] | Powder 298K [33] |
|---|---|---|---|
| Formula | CuInP$_2$S$_6$ | CuInP$_2$S$_6$ | CuInP$_2$S$_6$ |
| Crystal symmetry | Monoclinic | Monoclinic | Monoclinic |
| Space group | $Cc$ | $Cc$ | $Cc$ |
| Unit cell parameters | a = 6.0956 (4) Å | a = 6.0908(2) Å | a = 6.100 Å |
| | b = 10.5645(6) Å | b = 10.5615(3) Å | b = 10.557 Å |
| | c = 13.6230(8) Å | c = 13.1737(4) Å | c = 13.194 Å |
| | α = 90° | α = 90° | α = 90° |
| | β = 107.101(3)° | β = 99.1160(10)° | β = 98.52° |
| | γ = 90° | γ = 90° | γ = 90° |
| Volume (Å$^3$) | 838.5(1) | 836.73(4) | 841.17 |
| Z value | 4 | 4 | 4 |
| Density (mg/m$^3$) | 3.427 | 3.435 | |
| Absorption coefficient (mm$^{-1}$) | 7.017 | 7.086 | |
| Characterization | X-ray diffraction | X-ray diffraction | Synchrotron diffraction |
| Radiation | Mo K-L$_{2,3}$ (λ= 0.71073 Å) | Mo (λ= 0.71073 Å) | Photon source (λ= 0.117418 Å) |
| Refined factors | R$^1$ = 0.051 wR$^2$ = 0.056 | R$^1$ = 0.0238 wR$^2$ = 0.0578 | R$_p$ = 12.2 R$_{wp}$ =11.2 |

**Table 1** Refined structural information of CIPS single crystals and powder sample [33,41,42].

The out-of-plane ferroelectricity and ionic conductivity of CIPS originate from the spatial instability of the Cu$^I$ cation. Fundamentally, monovalent Cu$^I$ favors lower coordination instead of sitting at the center of the S octahedron, which is driven by a second-order Jahn-Teller coupling between the filled 3$d^{10}$ manifold and the empty 4$s$ orbital [43-45]. In their following work, Maisonneuve and co-workers revealed more structural details by looking into the thermal evolution of the Cu$^I$ cation in its sublattice [46]. As shown in **Fig. 2**, they found strong thermally-activated smearing of the Cu$^I$ electronic density along the layer normal, which can be accounted for by three partially filled crystallographic sites: (i) quasi-trigonal Cu1, shifted from the centers of the octahedra, (ii) octahedral Cu2, located in the centers of the octahedra, and (iii) nearly tetrahedral Cu3, which penetrates into the vdW gap [46]. In addition, a twofold axis doubles each Cu$^I$ sites: Cu$^u$ is shifted upward from the middle of the layer, and Cu$^d$ is shifted downwards from it. At 153 K, the Cu1$^u$ position located 1.55 Å above the layer midplane is almost completely filled, meaning that the quasi-trigonal Cu1 site is the ground state of the CIPS ferroelectric phase. As the temperature increases, the Cu1$^u$ occupancy decreases monotonically to 85% at 305 K, while the Cu1$^d$ site is filled in the meantime. Above $T_C$, the Cu1$^u$ and Cu1$^d$ sites become equivalent and the centrosymmetric structure (monoclinic space group *C2/m*) appears with twofold axis through the octahedral center. Such observations imply that hopping motions between the Cu1$^u$ and Cu1$^d$ and even between the intra- and interlayer sites already occur in the ferroelectric phase [46].

Recently, the studies by You et al. [40] and Zhou et al. [41] provide more insight into the fine structure of Cu$^I$ occupancy. Additionally, they found evident asymmetry distribution of the electronic density around the Cu1 site below the $T_c$ (**Fig. 2b&c**), rather than a harmonic vibration around the energy minimum position. This is a reminiscence of the high-temperature Cu3 site. Furthermore, first-principles calculations based on density functional theory (DFT) also indicates a metastable interlayer site with a free energy of merely 14 meV above the ground state [40]. As a result, the interlayer hopping of the Cu$^I$ cation can be easily excited thermally, leading to a broad spread of the occupancy distribution. The metastability of Cu$^I$ above and below the trigonal S plane resembles that found in zinc-blende CuCl, again due to the second-order or pseudo Jahn-Teller coupling [45]. The displacive instability of Cu$^I$ results in two intriguing properties of CIPS, namely, ferroelectricity (off-center ordering) and ionic conduction (long-range migration), which will be discussed in details in the following sections.

To maintain the structural stability, In$^{III}$ cation in the adjacent S octahedron has to displace oppositely to Cu$^I$, giving rise to a second polar sublattice. The consequent monolayer is a rare case of uncompensated ferrielectric ordering in a quasi-two-dimensional (2D) lattice. The next layer rotates 180 degree and shifts $\frac{1}{2}\vec{c}$ (or $\frac{1}{2}\vec{c}^* + \frac{1}{3}\vec{a}$) in reference to the first layer, in which the c* axis is perpendicular to the layer plane. As a result, the unit cell is monoclinic containing two crystal layers. The monoclinic angle is reflected by the horizontal shift of the stacking. It is worth noting that the discrepancy in the refined *β* angle is due to the different selections in terms of unit cell as depicted in **Fig. 1a**. Usually, smaller monoclinic angle is used in standard unit cell.

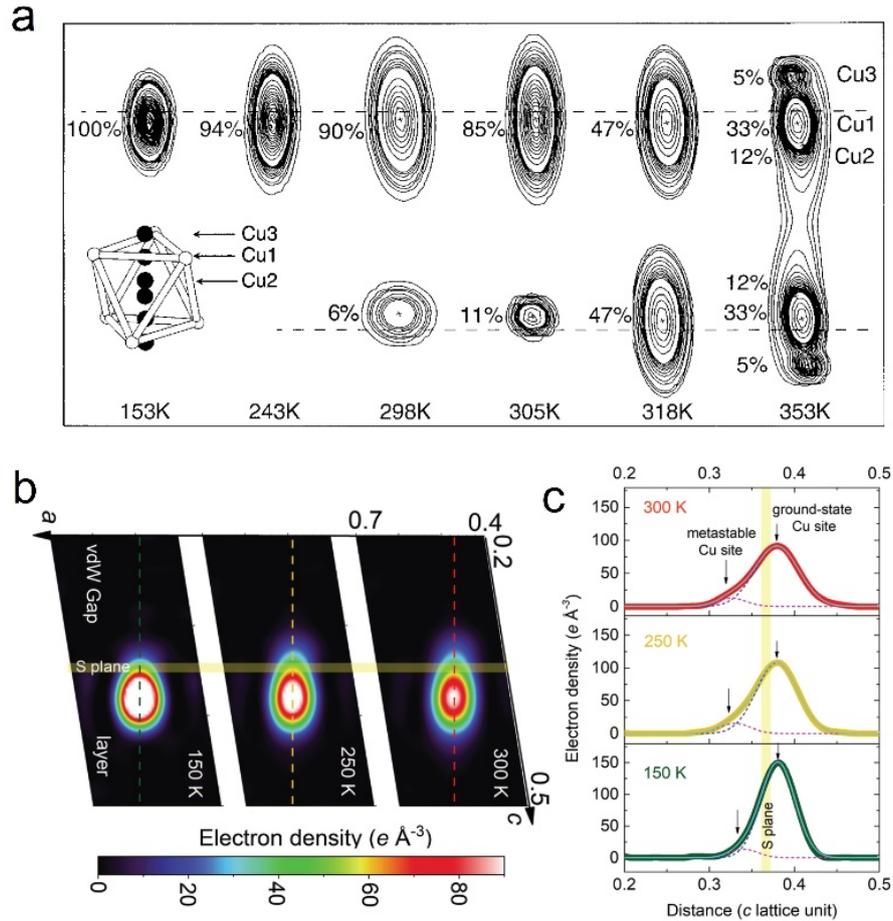

**Fig. 2 a,** Thermal evolution of the different copper site occupancies and the corresponding probability density contours in CIPS (main figure). Inset: A sulfur octahedral cage showing the various copper sites: the off-center Cu1, the almost central Cu2, and Cu3 in the interlayer space [46]. **b,** Temperature dependent electron density contour maps around the Cu sites. **c,** Corresponding line profiles of the electron density along the dashed lines shown in **b** [41]. The dash and solid lines are separate and combined peak fits using Gaussian function, respectively. Arrows denote the positions of Cu sites [41].

### 2.2. Temperature and compositional phase transitions

A first-order paraelectric (*C/2*) to ferroelectric (*Cc*) phase transition in CIPS was firstly revealed in 1994 by Simon et al. using calorimetry, X-ray powder diffraction and dielectric characterization (**Fig. 3**) [47]. From the differential scanning calorimetry (DSC) measurement, it was evident that the phase transition was accompanied by a latent heat and hysteretic behavior (**Fig. 3a**), both of which are the signatures of a first-order transition. Moreover, the dielectric response along the $c^*$ axis showed large anomaly at the transition temperature (**Fig. 3b&c**), typical behavior of a ferroelectric-paraelectric phase transition [47,41]. The first-order nature of the transition was again confirmed by the hysteretic behavior and the deviation from the Curie-Weiss law. Last but not least, temperature-dependent powder XRD was employed to monitor the structural changes across the phase transition, where all three lattice parameters exhibit obvious expansion (**Fig. 3d**) [46]. The behavior is in sharp contrast to that of conventional oxide ferroelectrics, which shows the shortening of the polar axis and contraction of the cell volume across the ferroelectric-paraelectric phase transition. In fact, it is an evidence of the negative longitudinal piezoelectric effect of CIPS, as will be discussed later. The synergetic study concluded with a $T_C$ of 315(5) $K$, which was verified by subsequent studies.

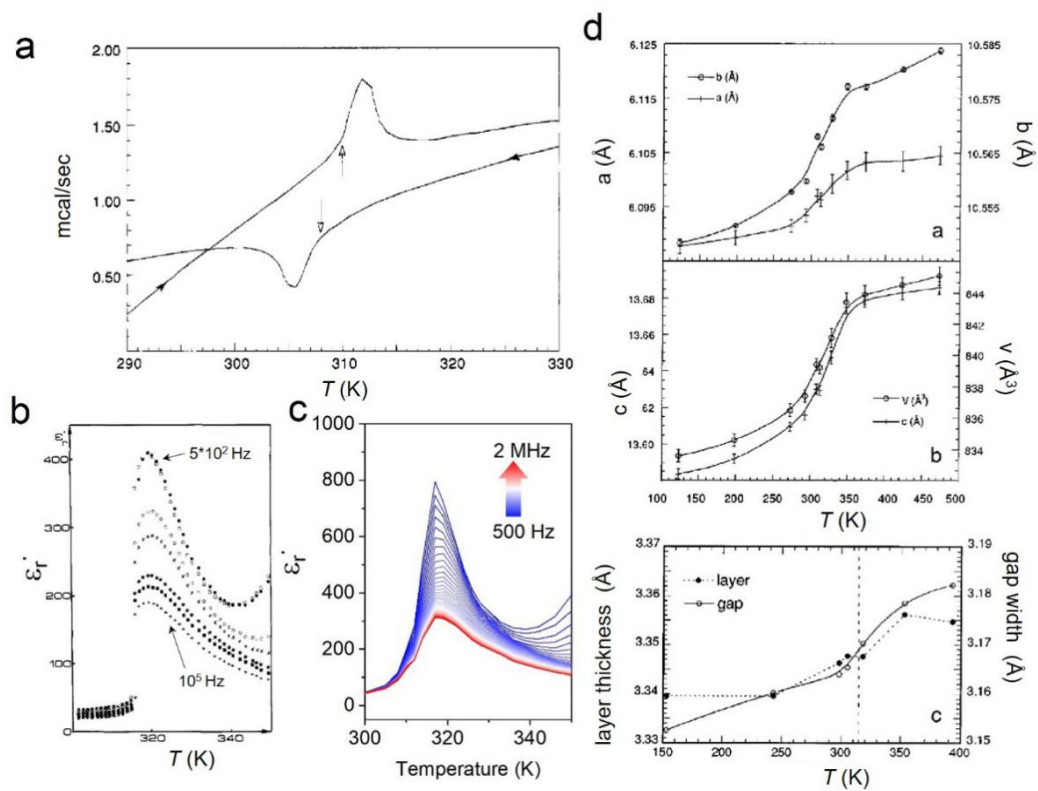

**Fig. 3 a,** DSC curves of CIPS powder along the warming and cooling cycles. **b** and **c**, Frequency-dependent dielectric anomaly of CIPS across the phase transition. **d**, Evolution of the lattice parameters of CIPS with temperature. [47,41,46]

The thermal evolution of $Cu^I$ cation suggests that above the phase transition temperature, the occupancy probabilities for $Cu1^u$ and $Cu1^d$ are equally 50% [46,47]. In other words, the copper is both spatially and temporally disordered in the paraelectric phase. Hence, the first-order phase transition is primarily an order-disorder type. Anisotropic electronic distribution of the $In^{III}$ was also found above $T_C$, implying the inherent coupling between the sublattice distortions.

Chemical alloying has been employed to tune the ferroelectric phase transition characteristics as well as to introduce additional order parameters, such as magnetism, in CIPS [48,49]. Sulphur was also reported to be replaced by the same chalcogen group element, selenium. The resulting compound, $CuInP_2Se_6$, shows a drastic reduction of the $T_C$ down to ~230 K [50]. Furthermore, a change in the stacking pattern leads to trigonal symmetry ($P$-31c → $P$31c) of the lattice instead of a monoclinic one. The $T_C$ reduction in the selenide compound can be attributed to the higher covalency and larger ionic radius of the Se anion, which relieves the potential constraint for Cu cation hopping compared to the sulfide analog. The mixed-anion intermediate compounds display a broad region of dipole glassy states in the phase diagram, with a possible morphotropic phase boundary (MPB) in between (**Fig. 4a**) [51]. Similarly, suppressions of the ferrielectric ordering were also found in cation substitution cases, including Cu replaced by Ag [52], and In replaced by Cr [48] or V [53]. Nevertheless, the introduction of Cr cation brings in the coexistence of antiferromagnetic and antiferroelectric orders, making $CuCrP_2S_6$ a potential multiferroic material (**Fig. 4b**) [48]. These phase transition behaviors under chemical alloying can be understood as the modulation of the Cu ordering potential relative to the thermal energy due to the delicate change in local chemical environment.

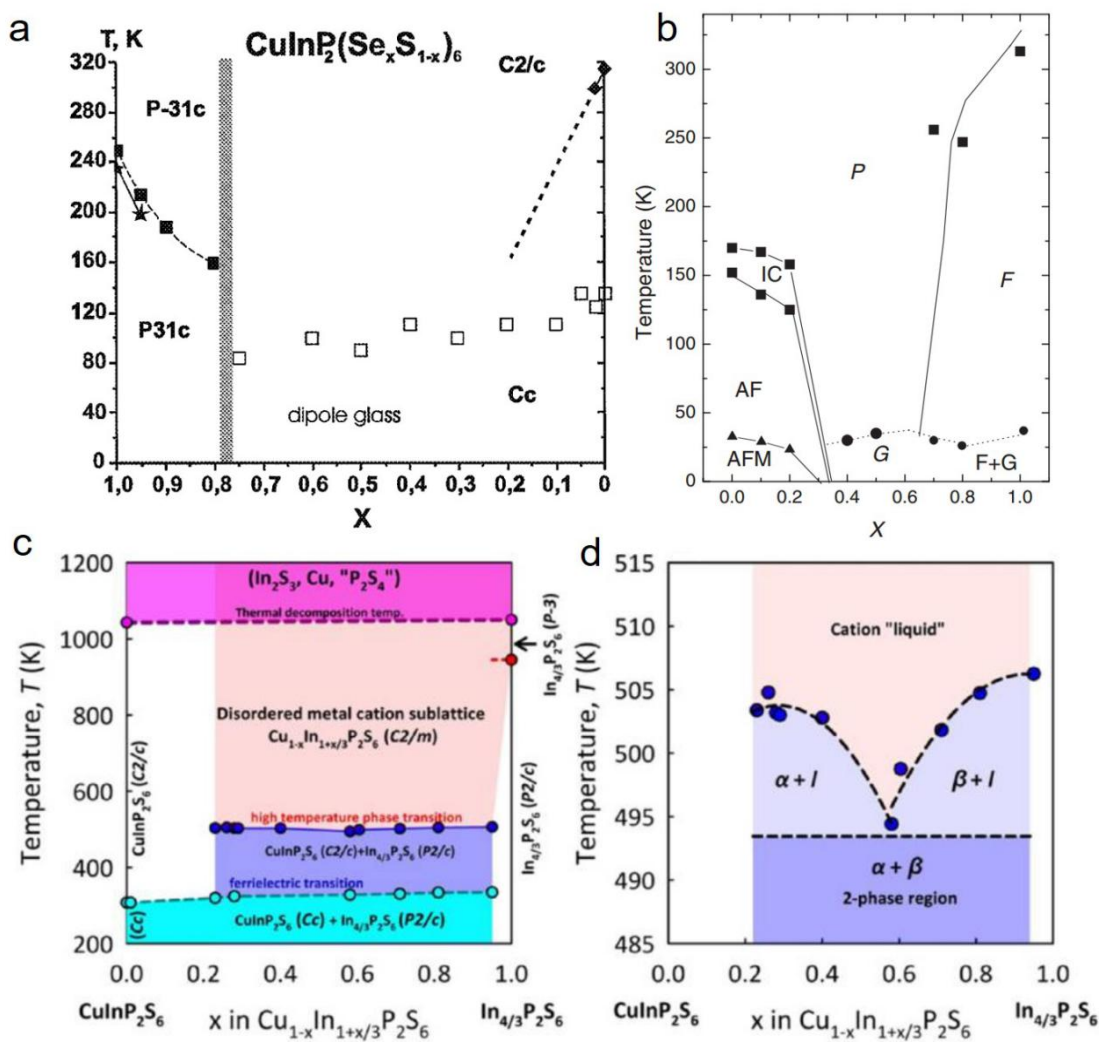

**Fig. 4 a-c,** Temperature and compositional phase diagrams of (a) $CuInP_2(Se_xS_{1-x})$ [51], (b) $Cu(In_xCr_{1-x})P_2S_6$ [48], and (c) $Cu_{1-x}In_{1+x/3}P_2S_6$ system [33]. **d**, Enlarged view of the eutectic phase transformation region of (c) [33].

There are, however, some other effective ways to enhance the ferroelectric order, such as hydrostatic pressure [54-56], In-rich structure [57,58] and chemical phase separation [38], which were shown to increase the $T_C$ to 360, 330, and 338 K, respectively. In 1998, Grzechnik et al. investigated the pressure-induced first-order phase transition from monoclinic to possibly trigonal structure at about 4.0 GPa based on Raman spectroscopy [54]. A few years later, Shusta et al, reported the pressure-dependent $T_C$ of CIPS by looking into the dielectric and optical properties [55]. Counterintuitively, increasing pressure causes a linear increase of the phase transition temperature with a coefficient of $\partial T_c/\partial p$ = 210 K/GPa, in stark contrast to conventional oxide ferroelectrics, which usually show decrease of $T_C$ under high pressure [59]. This, again, points to the anomalous longitudinal negative piezoelectricity of CIPS.

Interestingly, Dziaugys and coworkers found that excessive Indium addition could increase the phase transition temperature of CIPS from 315 to about 330 K [57,58]. The mechanism was later explained by Susner and coworkers, who discovered that off-stoichiometry Cu/In ratios will lead to phase mixtures containing CIPS and $In_{4/3}P_2S_6$ (IPS) [38]. IPS is non-polar with a monoclinic structure (space group $P2_1/c$). The layer spacing of the mixed-phase compound, defined as the perpendicular distance between the midpoints of neighboring layers, increased linearly from 6.5054(1) Å for IPS to 6.5146(2) Å for CIPS. As a result, the enhanced $T_C$ of CIPS in the mixtures can then be explained by the compressive chemical pressure imposed by the non-polar IPS phase, in analogue to the hydrostatic pressure cases. What is more intriguing is that they discovered the phase separation was achieved through a coherent spinodal decomposition at around 500 K, above which the cation sublattices melt into a cation-disorder liquid-like state [38]. A complete compositional phase diagram of the CIPS-IPS system is shown in **Fig. 4c-d** [33]. At low temperature, the compound is a mixture of the ferroelectric CIPS (*Cc*) phase and the paraelectric IPS ($P2_1/c$) phase. The former transforms into a paraelectric structure (*C2/m*) above the $T_C$ of CIPS. Across the sublattice melting temperature ($T_s$), the phase separation disappears and the compound undergoes cation sublattice melting into a single paraelectric phase $Cu_{1-x}In_{1+x/3}P_2S_6$ (*C2/m*), with disordered metal cations sublattice ($CuInP_2S_6$ (*C2/c*) + $In_{4/3}P_2S_6$ (*P2/c*) → $Cu_{1-x}In_{1+x/3}P_2S_6$ (*C2/m*)) [33]. Finally, the thermal decomposition takes place at ~1060 K.

### 3. Room temperature ferroelectricity

#### 3.1 Spontaneous polarization

The atomic origin of the spontaneous ferroelectric polarization ($P_s$) in CIPS is the off-center ordering of $Cu^I$ cations driven by the second-order Jahn-Teller effect, involving the hybridization of the localized $d^{10}$ states and the empty *s-p* orbitals that minimizes the total energy [43-45]. To balance the large structural distortion caused by the copper off-centering, $In^{III}$ cations in the adjacent sulfur cages have to displace in the opposite direction, resulting in a colinear ferrielectric lattice. As such, the total polarization can be calculated by summing up the products of the Born effective charge and relative atomic displacement of each individual ion, and divided by the cell volume. Maisonneuve and coworkers estimated the $P_s$ of CIPS to be around 3 $\mu C/cm^{-2}$ at room temperature, and their experimental result of 2.55 $\mu C/cm^{-2}$ is fairly close to it (**Fig. 5a-b**) [46]. The modern Berry-phase approach predicts a $P_s$ of 3.34 $\mu C/cm^{-2}$ at 0 K [40], in good agreement with the Born effective charge method [46]. However, Liu *et al.* from Nanyang Technological University (NTU), Singapore reported a $P_s$ of CIPS in the range of 3.5-4 $\mu C/cm^{-2}$ [13, 40, 41], which was also confirmed independently by another group [30]. The discrepancy was explained by the partial fill of the interlayer Cu3 site, which produces larger dipole moment [40]. Furthermore, the accurate determination of the polarization value is complicated by its anomalous switching behavior and considerable ionic conduction (leakage) in CIPS. For example, an as-grown crystal could show square yet unsaturated loops even after tens of repetitive cycles (**Fig. 5c**). This may explain the lower polarization value in the early reports.

For a long time, it was taken for granted that the ferroelectric polarization of CIPS is perpendicular to the layer plane with no in-plane polarization. This is seemingly true by looking at the off-center displacement of $Cu^I$ cation shown in **Fig 2.** However, a recent paper reported the study of in-plane polarization in CIPS [23]. Indeed, by scrutinizing the lobe of $Cu^I$ electron density, we can find it slightly tilted with regard to the layer-plane normal, which is more pronounced in the high-temperature case for the tetrahedral Cu3 site. The atomic origin for this tilt is the slight horizontal misalignment of the adjacent layer, leading to a canted interlayer Cu-S bond. The behavior is, however, consistent with the low symmetry (monoclinic $Cc$) of the lattice. Moreover, the authors of ref 23 also reported a structural transition from monoclinic to trigonal, namely, a change in the stacking pattern, below a critical thickness around 90-100 nm in CIPS. Finite size effect is common in 2D materials, but usually happens at few-layer limit. It is thus surprising that the critical thickness of CIPS is so large. The finding surely warrants further investigations.

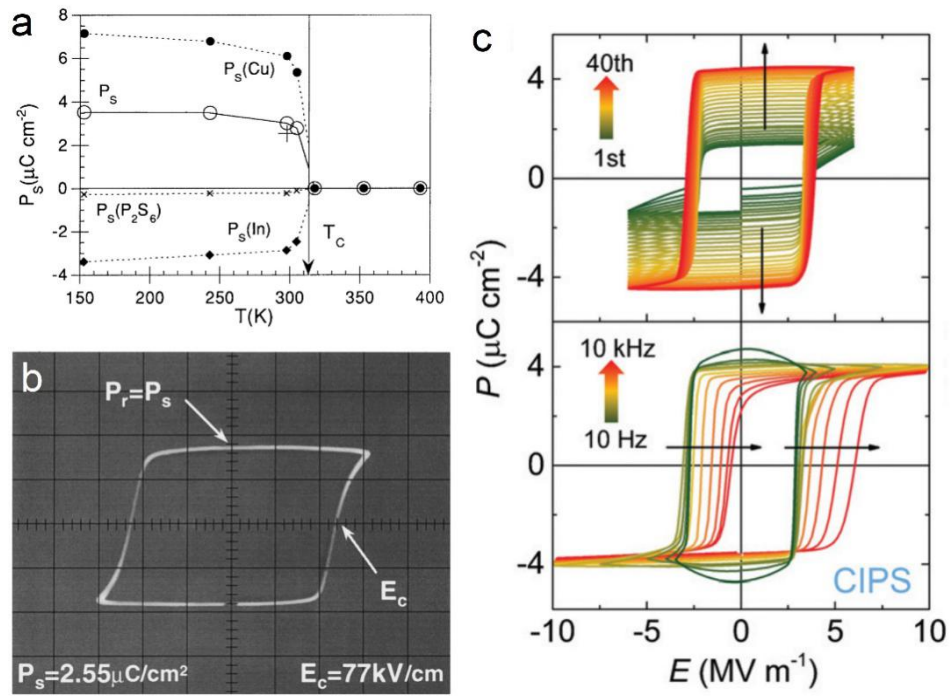

**Fig. 5 a,** Total polarization $P_s$ and sublattice contributions $P_s$ ($Cu^I$), $P_s$ ($In^{III}$), and $P_s$ ($P_2S_6$) calculated using Born effective charge and crystallographic results; the cross denotes the room-temperature experimental value [46]. **b,** Room-temperature saturated $P$-$E$ hysteresis loop of CIPS measured at 50 Hz [46]. **c,** Recent experimental polarization-electric field ($P$-$E$) loops of CIPS crystal; upper panel: cyclic hysteresis loops of an as-grown crystal under constant field and frequency; lower panel: frequency-dependent $P$-$E$ loops [41].

### 3.2 Dielectric relaxation and ionic conduction

Based on Jonscher's theorem on dielectric relaxation in solids [60,61], the low-frequency response is dominated by carrier polarization, while dipolar polarization relaxation following Debye behavior prevails at high frequency before phononic and quantum effect come into play. This also applies to CIPS, whose dielectric spectra were extensively studied to reveal its temperature-dependent dipole dynamics, especially across the phase transition [46,47, 52,62-64]. An overall survey of the dielectric spectra in both temperature and frequency domains are summarized in **Fig. 6**. In the low-frequency zone (DC limit to 1 MHz) (**Fig. 6a-b**), the spectra reflect mainly the carrier conduction characteristics, which are highly temperature dependent. Specifically in CIPS, the carrier polarization comes from ionic conduction as will be discussed in detail later on. Entering the high-frequency zone (10 MHz -1 GHz), strong dielectric dispersion begins to take place, which is most pronounced about the $T_C$ anomaly (**Fig. 6c-d**) [64]. Above 500 MHz, the phase transition peak is completely suppressed, consistent with a Debye-type dielectric relaxation of the dipolar polarization. The dispersion behavior is typical for order-disorder ferroelectrics [65,66], which are characterized by a single dispersive mode with a relaxation time much longer than the optical phonons. At low temperature, an additional weak dielectric anomaly occurs around 150 K, with apparent frequency dependence. This is in analogue to the phase transition in relaxor ferroelectrics. Hence, Dziaugys and coworkers attributed the dispersion to a dipole-glass transition [52]. Such a description is a bit misleading since the ferrielectric ordering persists across this transition point. A more plausible picture will be a complete freezing of $Cu^I$ cation from thermal hopping at this temperature as evidenced by the low-temperature electron density distribution. However, possible origin due to traces of impurity phase IPS can not be excluded either.

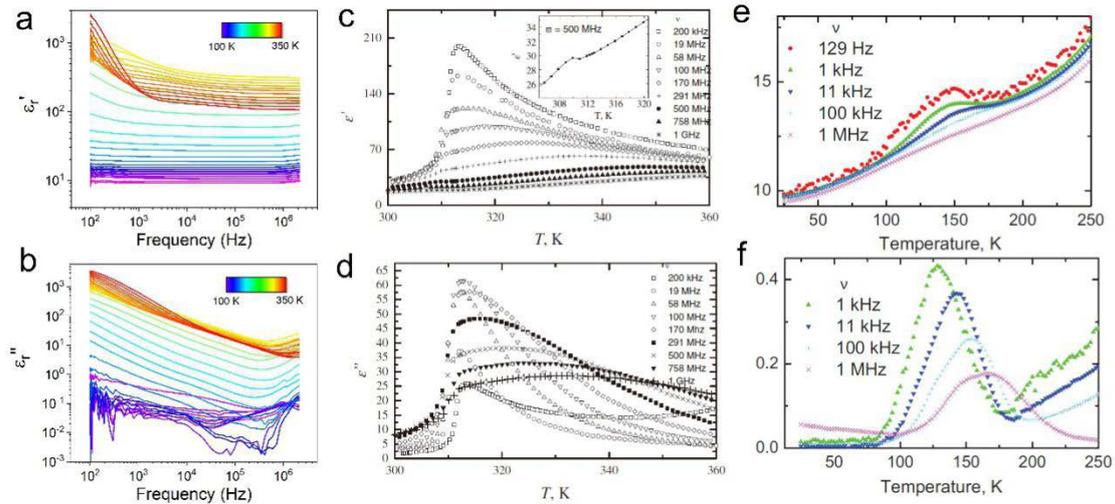

**Fig. 6 a-b**, Dielectric spectra of CIPS from $10^2$-$10^6$ Hz at various temperatures [41]. **c-d**, Temperature-dependent dielectric spectra of CIPS in high-frequency zone up to 1 GHz [64]. **e-f**, Low temperature dielectric anomaly of CIPS [52].

In terms of dielectric anisotropy, Dziaugys et al. compared the dielectric spectra measured parallel and perpendicular to the vdW crystal layers in both CIPS and $CuInP_2Se_6$ [50]. They found that dielectric anomaly across the phase transition was observed only in the direction perpendicular to the layers, which implies highly anisotropic inter- and intra-layer coupling. In contrast, the ionic conduction parallel to the layers was higher than that along the perpendicular direction, and the former also showed lower activation energy.

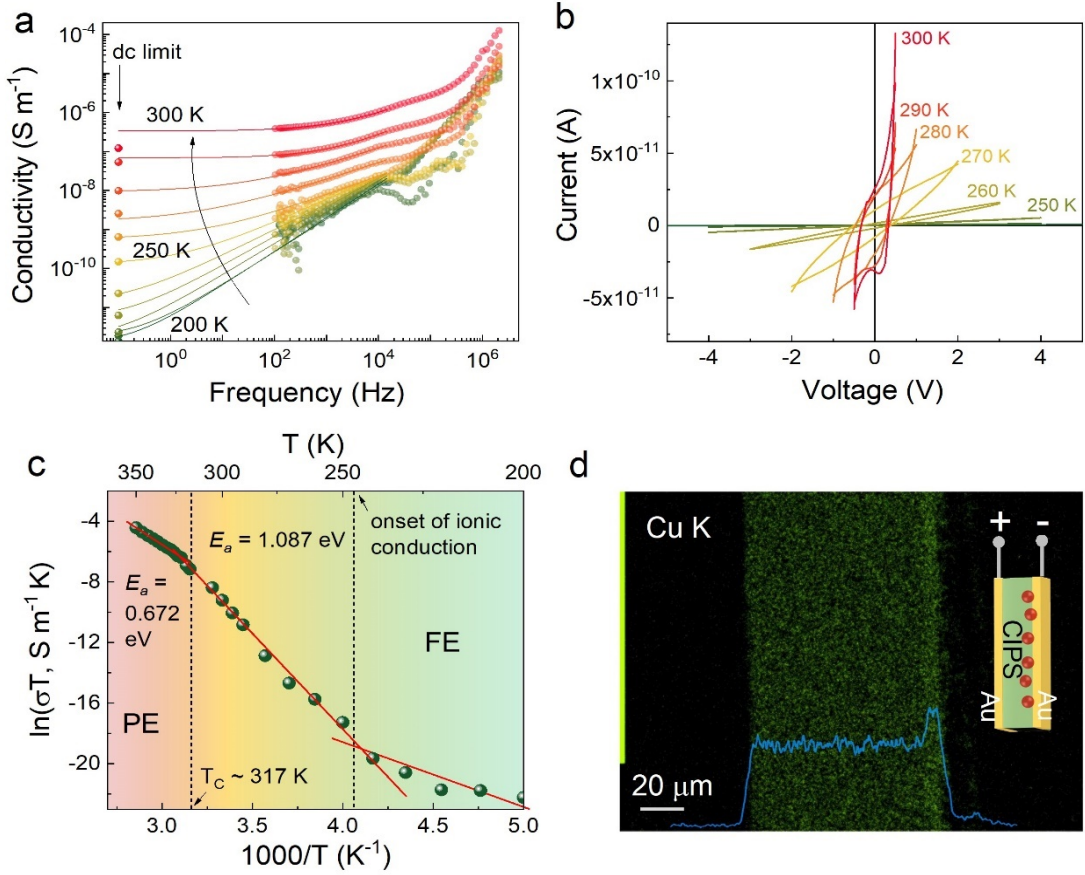

**Fig. 7 a,** Temperature-dependent AC conductivity of CIPS extrapolated to the DC limit. **b,** DC current-voltage loops under forward and reverse sweeps at various temperatures. **c,** Arrhenius plot of DC conductivity of CIPS. The solid lines are linear fits to the data. PE: paraelectric phase. FE: ferroelectric phase. **d,** EDX Cu map of a CIPS capacitor after DC stressing. The average horizontal line profile is overlaid on the map. The inset shows the device layout. Red circles denote the accumulation of Cu ions. [41]

Finite ionic conductivity was proposed in CIPS as early as the discovery of its ferroelectricity [46]. Arrhenius behaviors of the DC conductivity were found in both CIPS and $CuCrP_2S_6$ with comparable activation energies. Considering the strong Cu hopping motion seen in the structural results, $Cu^I$ migration was then concluded as the origin of the electrical conduction. This, however, is not necessarily true, given that electronic conduction can also be thermally activated, such as Poole-Frenkel effect. AC conductivity $\sigma$ is related to the imaginary dielectric permittivity $\varepsilon_r''$ through the following equation

$$\sigma = 2\pi f \varepsilon_0 \varepsilon_r''$$

Where $f$ is the frequency and $\varepsilon_0$ is the vacuum permittivity. It is then possible to extrapolate the conductivity to the DC limit by fitting to Jonscher's power law [61]

$$\sigma = \sigma_{dc} + A(2\pi f)^s$$

where $\sigma_{dc}$ is the conductivity at the DC limit, A is the preexponential constant, and $s$ is the power law exponent. The results agree well with the static conductivity deduced from the current-voltage measurements (**Fig. 7a-b**). The activation energy $E_a$ of the conduction could then be calculated using Arrhenius law (**Fig. 7c**)

$$\sigma T = \sigma_0 e^{-E_a/kT}.$$

Generally, the $E_a$ in the paraelectric phase is around 0.6-0.7 eV, which is consistent among different reports [32,41,52,64]. In the ferrielectric phase, the variation of $E_a$ is much larger, possibly due to the uncertainty in measuring the low conductivity of the sample.

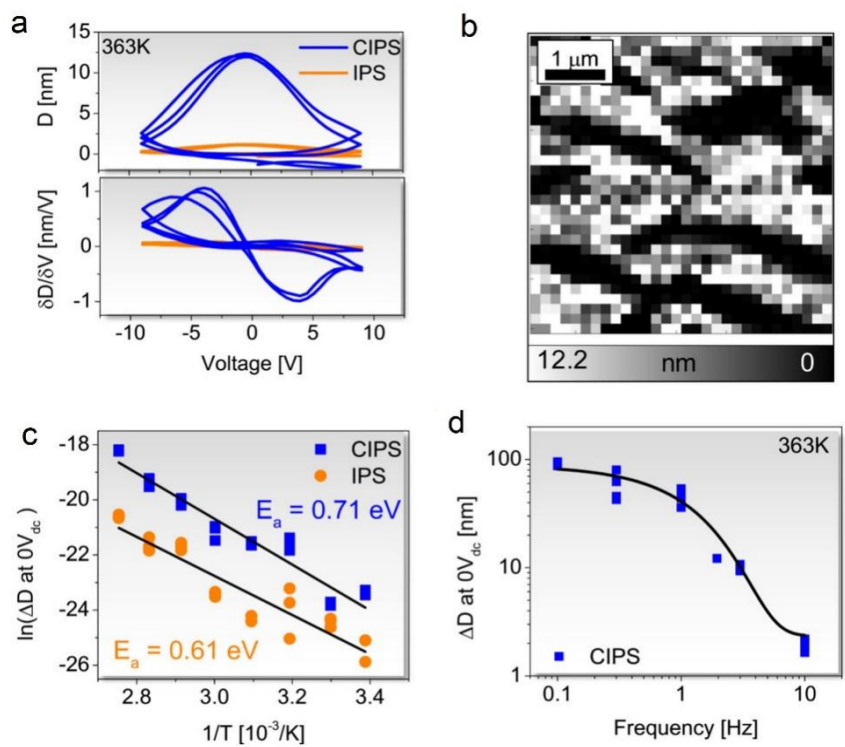

**Fig. 8 a,** Displacement $D$ and derivative of displacement $\delta D/\delta V$ at 363 K separated and averaged for CIPS and IPS phases as a function of applied DC voltage. **b,** Corresponding spatially resolved map of maximum displacement $\Delta D$. **c-d,** Analysis of reversible displacement $\Delta D$ at zero voltage for CIPS and IPS phases as a function of temperature (c) and DC voltage sweep frequency (d) [32].

Direct proof for the $Cu^I$ ionic conduction was demonstrated only recently. Zhou and coworkers performed compositional mapping on the cross section of the CIPS capacitor subjected to prolonged DC stressing using scanning electron microscopy with energy dispersive X-ray (SEM EDX) [41]. Accumulation of copper atoms was observed at the cathode interface. The local ionic process in CIPS was also investigated by Balke and coworkers, who employ scanning probe microscopy to detect and quantify ionic processes through the coupling between ionic migration and volume change [32]. Their measurements were conducted on mixed-phase crystals consisting of CIPS and vacancy ordered IPS phases. Large-scale ion-migration induced displacement was only observed in CIPS phase, but not in IPS phase, suggesting that the presence of Cu is required for large volume changes, as shown in **Fig. 8a-b**. The lattice displacement increases with increasing temperature and decreasing frequency (with a maximum value of ~90 nm observed at 0.1 Hz and 363 K), confirming the local ionic migration as the microscopic origin of the lattice expansion. As a thermally activated process, this reversible ion motion was featured by Arrhenius' law and the extracted activation energy in the paraelectric phase was 0.71 eV, consistent with the previous reports (**Fig. 8c-d**) [52,58,62]. Recently, the ionic migration effect of copper in CIPS was also reported by Xu and coauthors [67], and their high temperature measurements further enhanced the cation signals. Furthermore, an interesting ferroelectric switching phenomenon was observed that the in-plane electric field driven intralayer Cu hoping motions can induce the out-of-plane domain reversal [67].

### 3.3 Scaling effect

The surge of research on vdW layered materials is due to the possibility of obtaining mono- and few-layer samples and more importantly, the emerging phenomena and physics brought about. However, in ferroelectric materials, the imperfect screening of the spontaneous polarization results in a finite depolarization field that inversely scales with the thickness [68]. Approaching the 2D limit, the depolarization field can surmount the coercive field of the ferroelectrics, leading to either a multi-domain structure or reduced spontaneous polarization, both of which strongly suppress the switchable polarization of the ferroelectric material.

Belianinov and coworkers firstly reported the mechanical exfoliation of CIPS and studied the thickness dependent piezoelectric response by band excitation piezoresponse force microscopy (BE-PFM) [29]. They found at room temperature the flakes thicker than 100 nm showed bulk-like ferroelectric domain structures, whereas below 50 nm, the polarization vanished. Local polarization switching was demonstrated, however, not to a clean and deterministic level. Specifically, the piezoresponse amplitude is not uniform after switching and atypical switching hysteresis loops were observed. It appears that mixed-phase samples with IPS second phase were used and the "strange" loops could be explained by the electrochemical activity or the anomalous switching dynamics of CIPS as will be discussed in the following section. In the same year, the same group also reported room-temperature piezoresponse in a 20-nm-thick $Cu_{0.77}In_{1.12}P_2S_6$ flake [38]. One year later, they conducted more comprehensive investigation of the size effect and local switching behavior in CIPS thin flakes [69]. They found a progressive suppression of the piezoresponse for thicknesses ranging from 50 nm down to 10 nm, below which the polarization totally disappeared. Again, non-uniform and irregular switching hysteresis loops were shown, which were attributed to the irreversible electrochemical events at room temperature. Subsequent low-temperature measurements indeed produced more well-defined loops.

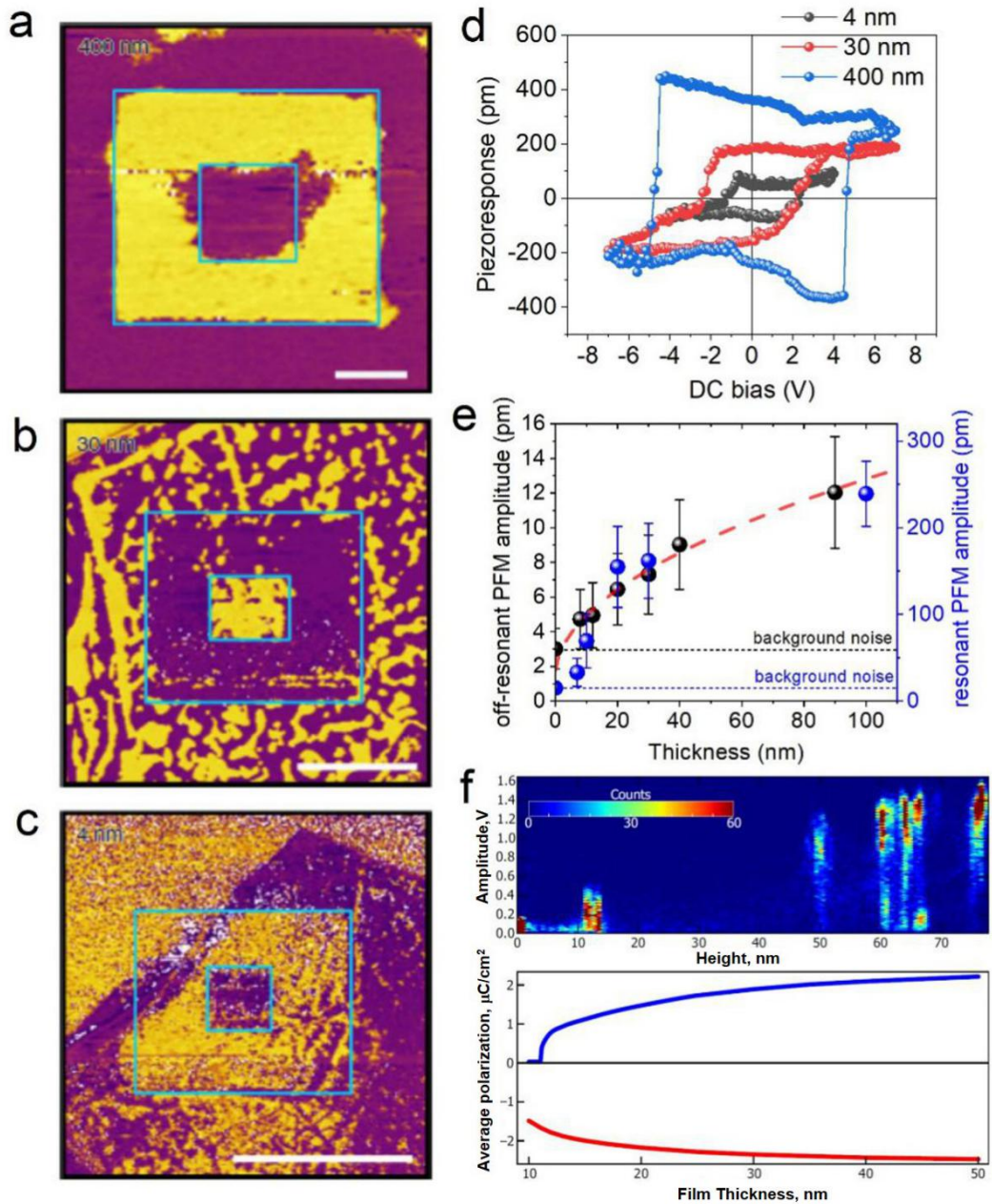

**Fig. 9 a-c**, PFM phase images for (a) 400 nm, (b) 30 nm and (c) 4 nm thick CIPS flakes after electric poling [13]. **d**, Local PFM hysteresis loops of different CIPS flakes. **e,** Evolution of Piezoresponse amplitude with flake thickness. **f,** Upper panel: 2D histogram plot of the PFM amplitude of CIPS flakes. Lower panel: Thickness dependence of calculated up- and down-polarizations. [69]

At about the same time, researcher from NTU also explored the possibility to maintain room-temperature ferroelectricity in CIPS down to the 2D limit. Liu and coworkers succeeded in the isolation of CIPS flakes down to bilayers [13]. Furthermore, they observed switchable piezoresponse in flakes as thin as 4 nm (six layers) using PFM, at least half of the minimal thickness reported previously (**Fig. 9a-d**). Certainly, the piezoresponse decreases with reduced thickness, especially below 50 nm (**Fig. 9e**). Summarizing these reports, the variation of the critical thickness in CIPS centers upon how to minimize the depolarization through sufficient screening. If CIPS is exfoliated onto an insulating substrate such as SiO$_2$, the critical thickness is around 50 nm. However, if doped or highly-doped Si is used, the limit can be pushed down to 10 nm or lower. In the work by Chyasnavichyus et al. from Oak Ridge National Lab (ONRL), Landau-Ginzburg-Devonshire (LGD) theory was applied to quantify the size effects by assuming a self-screening picture with finite carrier density [69]. However, the moderate bandgap of CIPS (~2.8 eV) precludes the electronic source of self-screening. A more plausible mechanism would be ionic screening via charged defects [70], on the grounds of the low activation energy for ionic conduction in CIPS. The charged defects, however, also act as random pinning centers, which explains the inhomogeneous switching behaviors of the samples previously shown. Furthermore, all the local switching studies exhibited more or less irreversible modifications of the sample surface after electrical poling, due to the considerable electrochemical activity under localized field at room temperature. The problem inevitably complicates the analyses of the local switching characteristics of CIPS.

Another mechanism to alleviate the depolarization field in thin flakes is the formation of domains. Since the in-plane polarization component is negligible, the domains formed in CIPS are purely 180° ferroelectric domains with no ferroelastic variants. In the absence of screening effects and defects, the domain scaling in ferroelectrics usually fits the Landau-Lifshitz-Kittel (LLK) law [71], where the domain width is proportional to the square root of the thickness. Thickness-dependent domain size evolution in CIPS has been observed in ref 13, which focuses more on the size effect (**Fig. 10a-c**). A quantitative study was performed recently by Chen and coworkers [34]. What they found is a deviation of the domain scaling from the classic LLK law, with an exponent about 0.65 instead of 0.5 (**Fig. 10e**). In fact, a rough estimation of the domain size in reference [13] gives rise to a scaling exponent of about 0.91 (**Fig. 10d**). However, the ferroelectric domains are mosaic-like with irregular shapes, making it difficult to calculate the size accurately. The random pinning of the domains by charged defect and asymmetric screening (conductive substrate and open surface) may also cause the breakdown of classic scaling law.

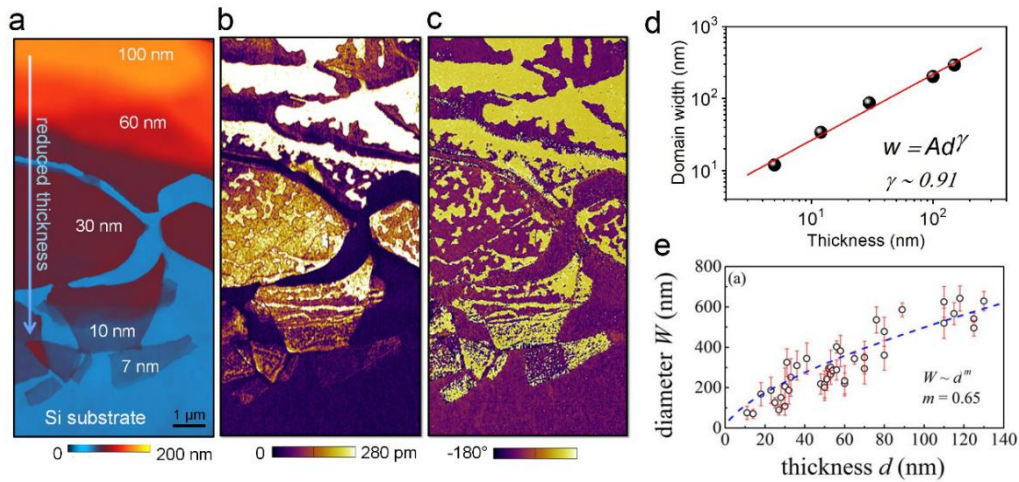

**Fig. 10 a,** Topographic, **b,** PFM amplitude, and **c,** PFM phase images of CIPS flakes with different thickness [13]. **d,** Scaling of the domain width with respect to flake thickness as derived from (a-c). **e,** Domain scaling results from another study [34].

Interesting domain scaling was observed in mixed CIPS-IPS compounds, as reported by the group in Oak Ridge National Laboratory (ORNL) [38]. Here, the domains refer to different crystal phases (either CIPS or IPS), whose volumetric percentages in the crystal are determined by the Cu/In ratios. For a given composition, it is possible to tune the domain size of the minor phase via the cooling rate. Apparently, the phase separation is still nucleation-limited and accompanied by the first-order structural transition, thus nullifying the previously proposed spinodal decomposition mechanism [33]. As shown in **Fig. 11**, the domain size is inversely proportional to the sample cooling rate. At slow rate, regular large-size domains are formed, while at fast cooling rate, large domain clusters comprised of multiple small domains are observed, which is reminiscent of the dendritic crystallization when the growth rate is limited by ion diffusion rate. The smaller domains around the large nucleate clusters also mimic the nanodomain ejected from the wavy domain walls under high electric field [72]. By controlling the domain size in the heterostructure system, it is possible to tune the ferrielectric characteristics of CIPS through a complex interplay between confined size effect and heterointerfacial strain effect, which deserves more study.

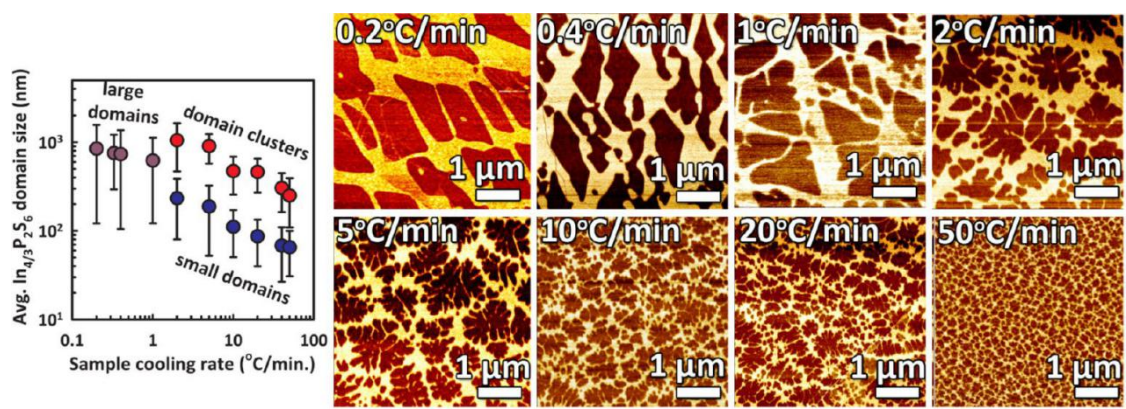

**Fig. 11** The "domain" evolution of the mixed-phase $Cu_{0.42}In_{1.23}P_2S_6$ compound cooled from above the cation disorder temperature at different rates. The left panel shows the corresponding "domain" scaling with respect to the cooling rate. [33]

### 3.4 Anomalous switching dynamics and roles of Cu cation

For ferroelectric materials, it is extremely important to understand the polarization switching dynamics which directly affects the operation speed and performance of ferroelectric devices [73,74]. Several models have been proposed to explain different ferroelectric switching behaviors, among which the two most classical models are the Kolmogorov-Avrami-Ishibashi (KAI) and nucleation-limited-switching (NLS) ones [75-78]. The former is based on the assumption of homogeneous nucleation and unrestricted domain growth, which is usually used to describe uniformly polarized single crystals and epitaxial films, for instance, poly(vinylidene fluoride-trifluoroethylene) copolymer (P(VDF-TrFE)) film and epitaxial $Pb(Zr_{0.4}Ti_{0.6})O_3$ (PZT) [73,75]. The latter describes region-by-region nucleation and switching, which applies to inhomogeneous systems with wide distribution of switching times, e.g. (Pb, La)(Zr, Ti)$O_3$ film, molecular ferroelectricity triglycine sulfate (TGS) single crystal [77,78].

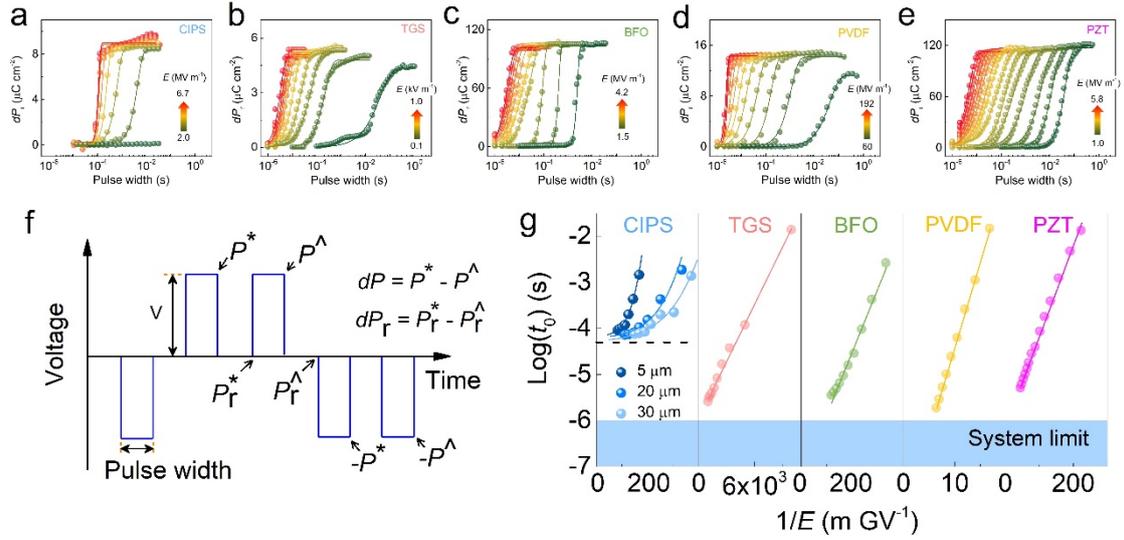

**Fig. 12 a-e,** Polarization switching spectroscopic curves of (a) CIPS, (b) TGS, (c) BFO, (d) PVDF, and (e) PZT fitted by KAI model. **f,** Voltage waveform used and the various polarization components measured in PUND measurements. **g,** Characteristics switching time $\log(t_0)$ versus $1/E$ curves for different ferroelectric materials, fitted by the Mertz' law, except for CIPS. [41]

The switching behavior of vdW ferroelectric CIPS was recently investigated by Zhou and coworkers [41]. For comparison, CIPS crystal and four other typical ferroelectrics were measured by traditional positive-up-negative-down (PUND) method, and fitted by the classical KAI model to obtain corresponding characteristic times, as shown in **Fig. 12g**. TGS, BFO, PVDF and PZT all revealed good linear relationship between the fitted switching time $t_0$ and $1/E$, in agreement with the empirical Merz's law [79]. However, all of the CIPS samples showed a strong deviation from the linear relationship, and not in line with any previously proposed models. Combined with the deconvoluted information of PUND and dielectric dispersions in both frequency and temperature domains, it is found that the polarization switching in CIPS is ionic-conduction-limited due to the strong electrostatic interaction between ferroelectric polarization and ionic defect dipoles [41]. Specifically, at low temperature when the ionic conduction is inactive, the ferroelectric polarization is fully pinned by the pre-existing defect dipoles. The situation persists until the onset of ionic conductivity at around 250 K, above which the defect-dipole moments become switchable, thus depinning the ferroelectric polarization. However, there is a mismatch between the switching speeds of the ferroelectric polarization and defect-dipoles, leading to a bifurcated kinetics (**Fig. 13a-b**). Hence, the switching speed is always limited by that of the defect-dipole reversal, which is tied to the slow ionic conduction activities.

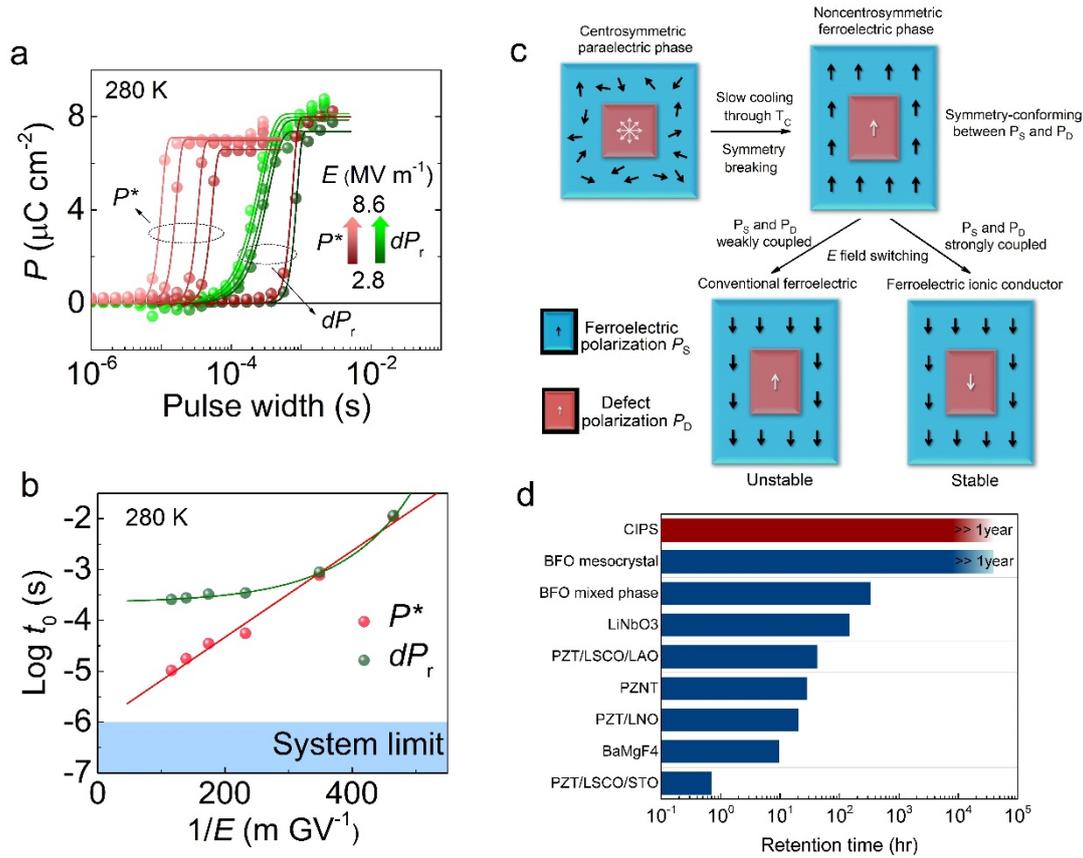

**Fig. 13 a,** Switching spectroscopic curves of total polarization $P^*$ (red) and remnant polarization $dP_r$ (green) at 280 K, fitted by the KAI model. **b,** Mertz' plot of the characteristic switching time $t_0$ versus $1/E$ for $P^*$ and $dP_r$ derived from (a). Linear fit is obtained only for $P^*$. **c,** Schematic illustration of the interplay between ferroelectric and defect polarizations in both conventional ferroelectrics and ferroelectric ionic conductors. Note that the volume of the defect region is exaggerated. In reality, a single defect dipole pins a lot more ferroelectric dipoles. **d,** Comparison of the retention time reported for different ferroelectric materials. [41]

The unusual polarization switching behavior is understood by the dual roles of Cu cations in CIPS. As elaborated in previous sections, both the ferroelectric polarization and ionic conduction arise from the displacive instability of the Cu cations through pseudo (second-order) Jahn-Teller effect. However, the two aspects appear mutually exclusive to each other, with the former featuring ordered parameter, whereas the latter requires structural disorder. This results in a paradox: how does the external electric field determine whether polarization switching or ionic conduction would occur? The difference lies in the activation energy, which, in turn, affects its characteristic time (**Fig. 13a-b**). For example, the activation energy for polarization switching in $BiFeO_3$ is about 0.4 eV [80]. Keep in mind that CIPS has much smaller spontaneous polarization and lower $T_c$. Furthermore, the actual energy required for polarization switching is usually much smaller than the theoretically predicted one due to the inevitable extrinsic defects and domain formation, which is referred as Landauer paradox in ferroelectric switching [81]. Thus, the activation energy for polarization switching in CIPS would be expected to be even smaller. In contrast, the activation energy for ionic conductivity is always above 0.7 eV based on literature reports [32,52,62]. It is, then, still possible to resolve these two events at different time scales in ferroelectric ionic conductor such as CIPS. However, they would become more and more convoluted at elevating temperature. Although large-scale ionic migration is detrimental to ferroelectric order, switchable defect-dipole at small quantity is surprisingly beneficial to polarization stability. Polarization retention of various ferroelectric materials was carried out by PFM characterization, and CIPS stands out from the rest with a domain relaxation time beyond one year (**Fig. 13d**). The permanent retention originates from the interplay between ferroelectric polarization and defect dipoles, as shown in **Fig. 13c**. Upon electrical switching, the defect-dipoles are aligned to the direction that stabilizes the matrix polarization due to its symmetry-conforming nature. When the voltage is removed, the internal field generated by the defect-dipoles will serve as a guard to prevent the polarization from back switching. The switchable defect-dipole interlock mechanism provides new strategy for permanent information storage.

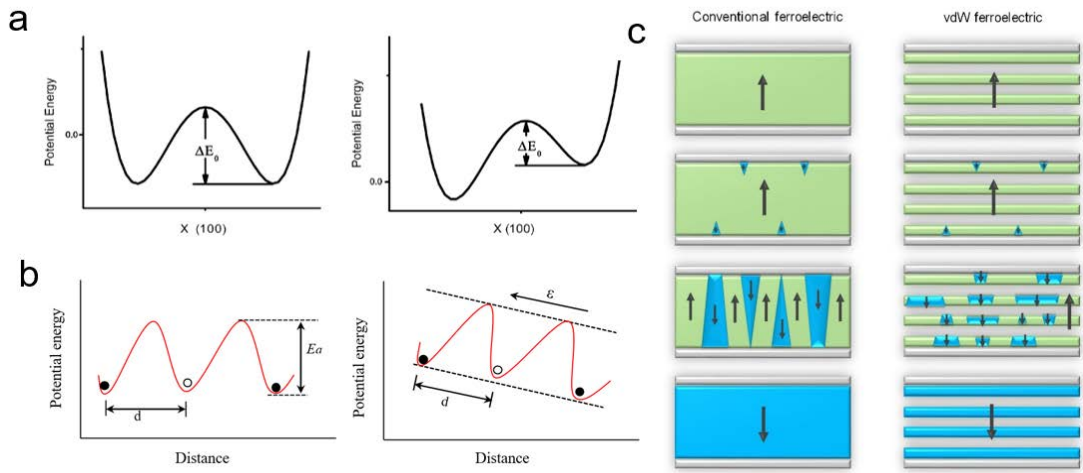

**Fig. 14 a-b,** Schematic comparison between (a) ferroelectric polarization switching [82] and (b) ionic conduction under external field. **c,** Schematic comparison of the domain growth processes between conventional and vdW-type ferroelectrics.

Last but not the least, we would like to highlight a possible difference in domain growth kinetics between conventional and vdW-type ferroelectrics (**Fig. 14c**). A classical description of the domain growth consists of nucleation, forward growth and sideway growth stages. Among the three stages, the forward growth is usually fast, which is limited by the speed of sound (~2500 m/s) [22]. For a 1-micron meter-thick sample, it takes less than 1 ns. However, in vdW-type ferroelectric, this process may be disrupted by the discontinuity of the lattice due to the vdW gaps. Combined with the abundance of charged defects, the intermediate stage would be comprised of complicated domain structures with prevalent charged domain walls, which accounts for a large window in the switching time scale. Collectively, the ferroelectric capacitor will appear to be piezoelectric inactive during this intermediate stage, which might explain the "strange" behaviors observed in local PFM measurements [69].

### 3.5 Negative piezoelectric effect

Piezoelectric effect, the interconversion between mechanical and electrical energy, is one of the most important and useful properties of ferroelectrics. Almost all ferroelectric materials have positive longitudinal piezoelectric coefficients ($d_{33}$), meaning that the lattice will expand when the applied electric field is along the existing polarization direction. The only previously known exception is ferroelectric polymer PVDF, which possesses negative longitudinal electrostriction and piezoelectricity. Although known for decades, the mechanism for its negative piezoelectricity has been under debates [83,84]. There has been no experimental confirmation of negative longitudinal piezoelectricity in other material system, despite some theoretical predictions [85,86]. As early as 2016, Liu and coworkers have pointed out that CIPS could possibly have a negative $d_{33}$ as that of PVDF [13]. The subsequent work by You and coworkers further confirmed this result and unraveled the microscopic origin of the giant negative piezoelectricity in CIPS [40]. By quasi-static strain-electric field (*S-E*) and dynamic small-signal $d_{33}$ measurements, they quantified the related material parameters of three prototypical ferroelectric materials, namely, PVDF, CIPS and PZT (**Fig. 15** & **Table 2**). The obtained $d_{33}$ and longitudinal electrostriction coefficient ($Q_{33}$) for CIPS is -95 pm/V and -3.4 m$^4$/C$^2$ at room temperature, both of which are among the highest in single-phase ferroelectrics.

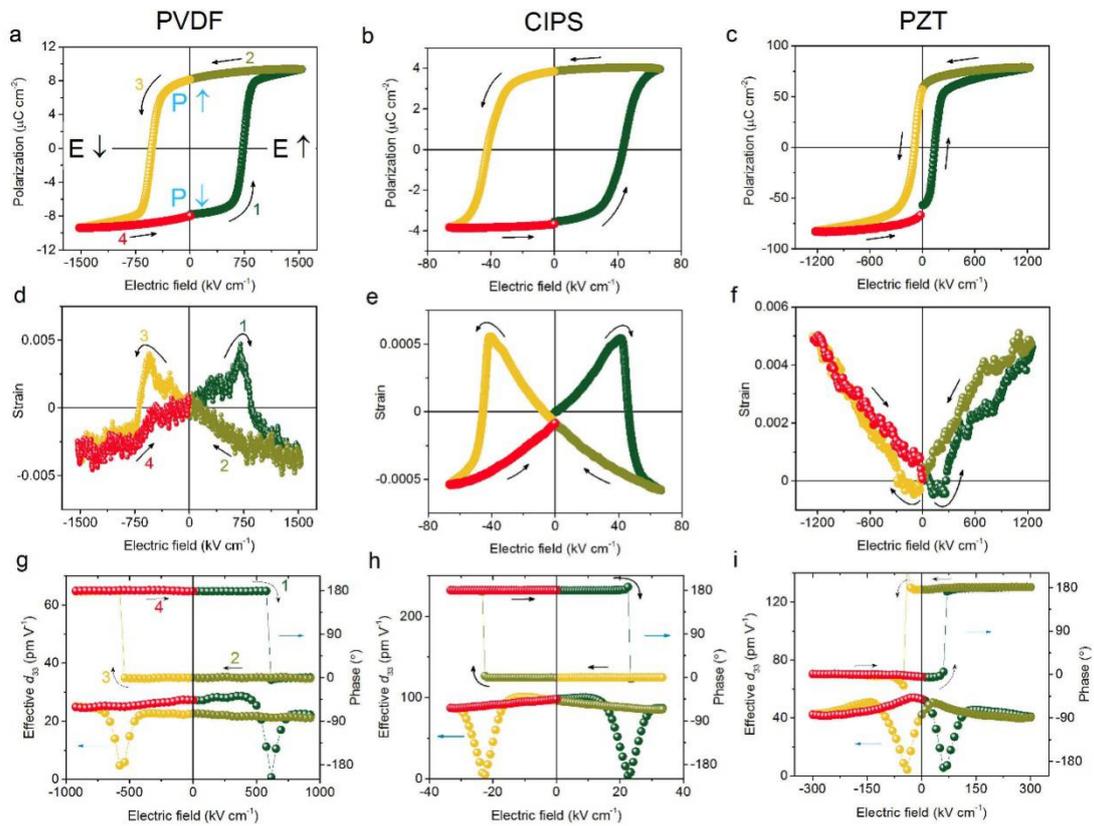

**Fig. 15 a-c,** Polarization–electric field (*P-E*) hysteresis loops of (a) PVDF, (b) CIPS and (c) PZT. **d-f,** Corresponding Strain-electric field (*S-E*) hysteresis loops of (d) PVDF, (e) CIPS and (f) PZT. **g-i,** ac small-field hysteresis loops of effective $d_{33}$ (amplitude) and phase signals for (g) PVDF, (h) CIPS and (i) PZT. [40]

| Material | $P_s$ (μC/cm$^2$) | $\varepsilon_{33}$ (at 10 kHz) | $d_{33}$ (pm/V) (at 10 kHz) | $Q_{33}$ (m$^4$/C$^2$) | $E_b$ (kJ/mol) | $C_{33}$ (GPa) | Coupling factor $k_{33}$ |
|---|---|---|---|---|---|---|---|
| PVDF-TrFE (70/30) | 8 | 8.2 | −25 | −2.2 | <20 (290–495) | 1–3 | 0.1–0.16 |
| CuInP$_2$S$_6$ | 4 | 40 | −95 | −3.4 | <20 (280–440) | 20–30 | 0.71–0.88 |
| Pb(Zr$_{0.4}$Ti$_{0.6}$)O$_3$ | 60 | 170 | 50 | 0.02 | 370–800 | 100–150 | 0.41–0.5 |

**Table 2** A list of extracted materials parameters. $P_s$, spontaneous polarization; $\varepsilon_{33}$, relative permittivity; $d_{33}$, longitudinal piezoelectric coefficient; $Q_{33}$, longitudinal electrostriction coefficient; $E_b$, intermolecular bond energy; $C_{33}$, Young's modulus. The numbers in the parentheses are the intramolecular bond energies [40].

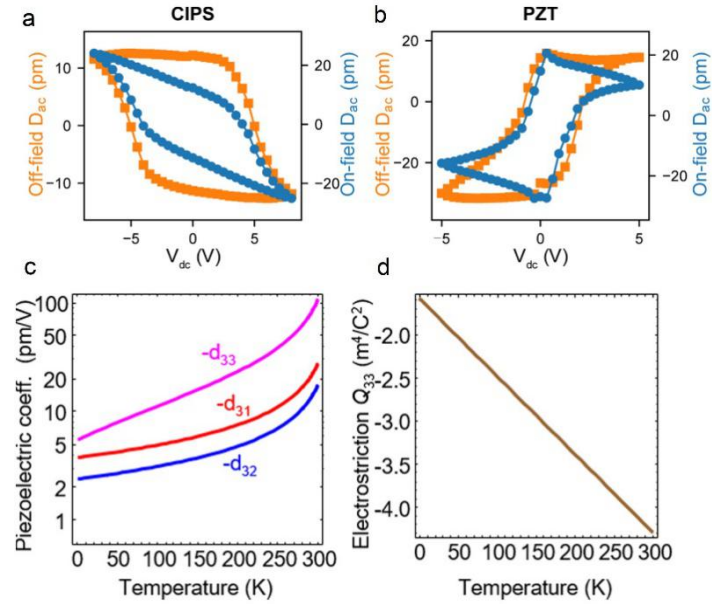

**Fig. 16 a-b,** Local switching displacement hysteresis loops obtained by PFM for (a) CIPS and (b) PZT. **c-d,** Calculated temperature-dependent (c) piezoelectric coefficients and (d) electrostriction coefficient $Q_{33}$. [37]

In the meantime, Neumayer and coworkers from ORNL provided more evidence of the negative electrostriction and piezoelectricity in CIPS using PFM, synchrotron X-ray diffraction, and phenomenological calculations [37]. As shown in **Fig. 16a-b**, CIPS and PZT exhibit completely opposite local displacement hysteresis loops, consistent with previous reports. Moreover, the electrostriction coefficients $Q_{13}$, $Q_{23}$, and $Q_{33}$ were extracted from the measured lattice constants across the $T_c$, by considering the contribution from both thermal expansion and spontaneous strain arising from ferroelectric polarization. Subsequently, the piezoelectric coefficients of bulk CIPS were calculated following $d_{3j}= 2\varepsilon_0\varepsilon_{33}Q_{j3}P_3$, as summarized in **Fig. 16c-d**. Near the ferroelectric Curie point at room temperature, the longitudinal piezoelectric coefficient is as high as -85 pm/V and the $Q_{33}$ up to -4.2 m$^4$/C$^2$, about two orders of magnitude higher than that of PZT.

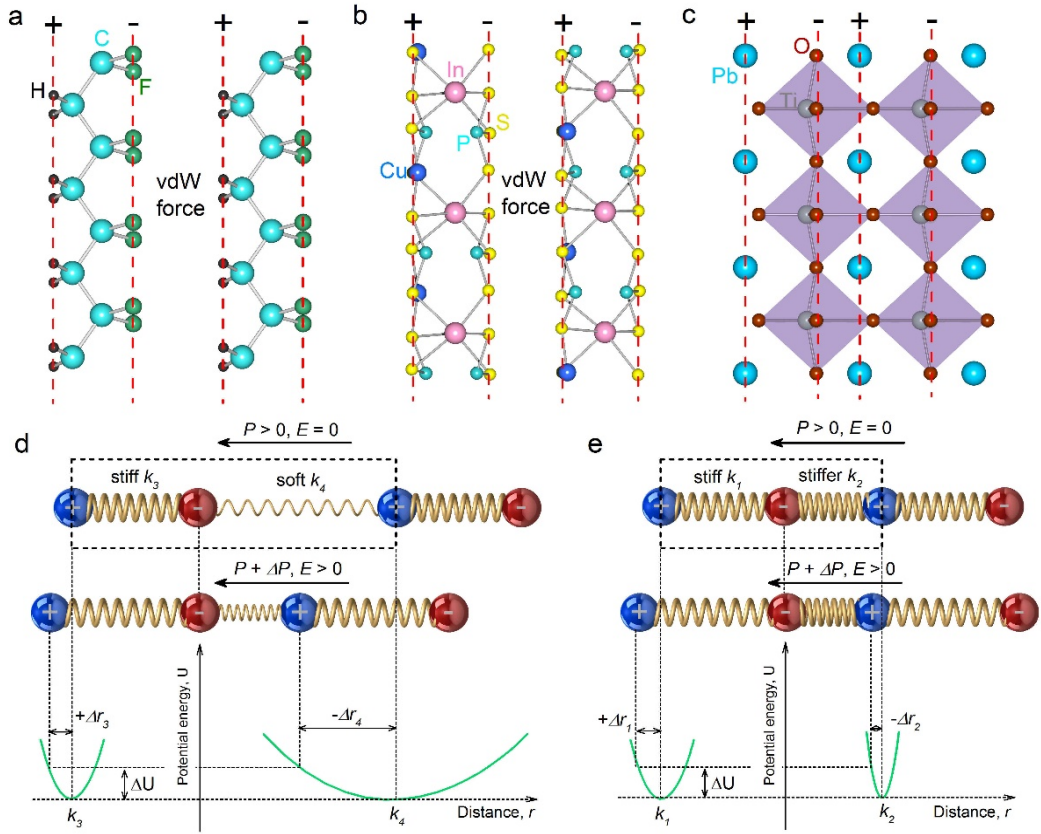

**Fig. 17** Simplified rigid ion model for piezoelectricity in polar solids. **a-c**, Correlations between crystal structures and dipole charges for (a) PVDF, (b) CIPS, and (c) PZT. **d,** Negative piezoelectric effect in polar solid with discontinuous (broken) lattice. **e,** Positive piezoelectric effect in polar solid with continuous lattice. The unit cell of the lattice is indicated by the dashed box. The polarization and electric field directions are denoted by the arrows. The lower parts show the pair potential energy profiles of the corresponding chemical bonds. The negative ions are taken as a reference point for the relative changes of bond lengths. [40]

In view of the lattice-dipole relationship in CIPS, PVDF and PZT, You and coworkers proposed a general mechanism based on the rigid ion model to explain the negative longitudinal piezoelectricity [40], which is also compatible with the previous "dimensional" model and Maxwell strain explanation [87,88]. The dipoles in PVDF and CIPS emerge within isolated layers/chains, which are held together through weak vdW interactions to form "broken" lattices and "quantized" electric dipoles (**Fig. 17a-b**). In comparison, conventional perovskite oxide 3D ferroelectrics carry electric dipoles in a continuous lattice with strong covalent/ionic bonds (**Fig. 17c**). The positive piezoelectric effect can then be understood since the expansion is always easier than the compression under the external electric field ($k_2 > k_1$), owing to the bond anharmonicity resulted from the preexisting spontaneous polarization. Whereas in the case of PVDF and CIPS, the large anisotropy between the strong intramolecular bonds and the weak intermolecular bonds results in the abnormal electromechanical coupling behavior. Specifically, when the external electric field is along the spontaneous polarization direction, the enhanced interlayer dipole-dipole interaction will dominate the lattice deformation, causing the vdW gap to shrink more than the expansion of the intramolecular bond ($k_3 \gg k_4$). To shed more light on the giant negative piezoelectricity from atomistic level, they also carried out detailed single crystal x-ray crystallography and DFT calculations. Both the experimental and theoretical results confirm that such giant negative longitudinal piezoelectricity and electrostriction result from the large displacive instability of Cu ions coupled with its reduced lattice dimensionality. The proposed mechanism thus implies the prevalence of negative piezoelectricity in ferroelectrics with low-dimension lattice, as what is indeed emerging in the recent literature [89,90].

## 4. Electronic structure and optical properties

### 4.1 Electronic structure

Although the off-center distribution of Cu in CIPS has been attributed to the pseudo Jahn-Teller coupling ever since its early studies [46]. It comes more like a posteriori based on other related systems from the stereochemical viewpoint [45]. Bercha and coworkers conducted the first detailed discussion on the origin of the spontaneous polarization and phase transition by combining group theory analysis and DFT-based *ab initio* calculations [91,92]. Starting from a high-symmetry trigonal proto structure, they correlated the elementary energy band splitting with the symmetry lowing of the lattice towards the paraelectric phase. In paraelectric phase, the valence band maximum consists of four weakly split subbands formed mainly by *d*-states of Cu (**Fig. 18a**). The non-degenerated electronic states near the band edge enables the pseudo Jahn-Teller effect through the vibronic coupling. When it transforms into the ferrielectric phase, the four states become more distant in energy and the corresponding four branches become more spread, indicating that the connection between the respective branches is weaker, as shown in **Fig. 18b**.

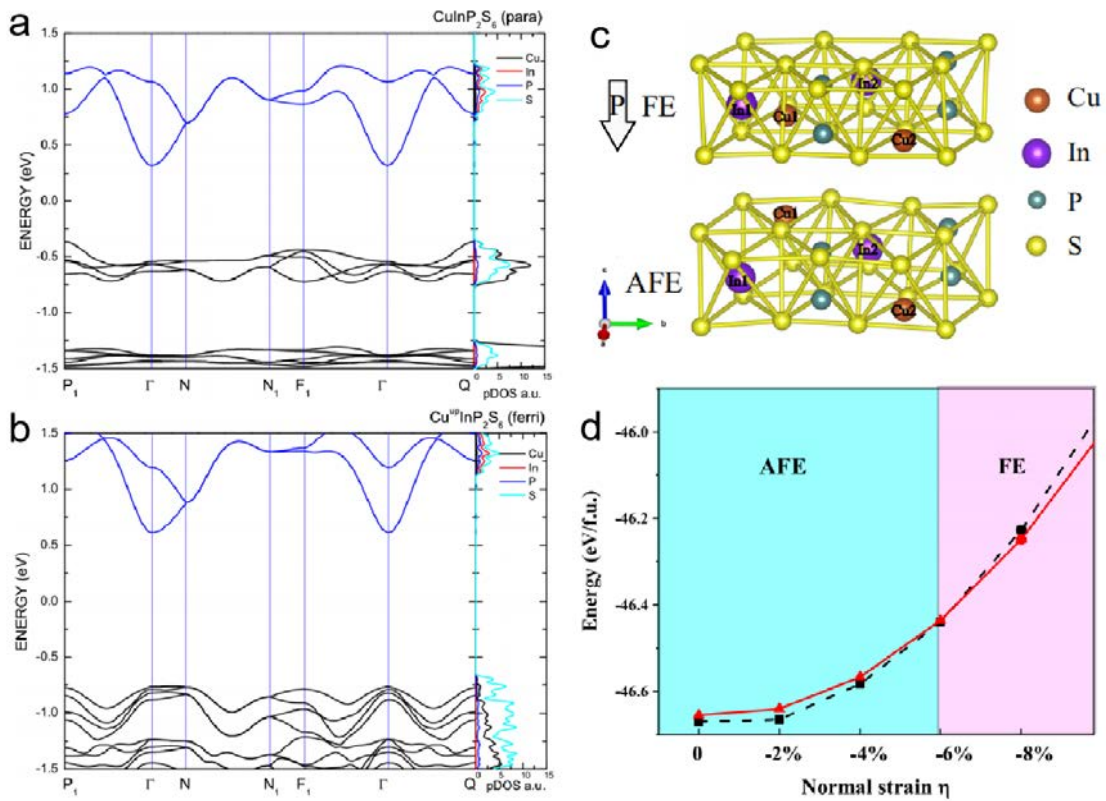

**Fig. 18 a-b,** Density of states for para-CIPS (a) and ferri-CIPS (b). **c**, Illustration of CIPS monolayer structure in FE state and AFE state. **d**, Total energy versus normal strain, where solid

line corresponds to FE structure, and dashed line stands for AFE structure. [92,39]

In 2019, Babuka et al. investigated the electronic band structure of monolayer CIPS across the phase transition, and the method of plane-wave pseudo-potential methodology within DFT was used for the first time [93]. The calculated projected density of states showed that the ferrielectric phase of CIPS has a direct bandgap ($E_g$=1.503 eV, $\Gamma$-$\Gamma$ transition), while paraelectric phase exhibited indirect bandgap (L-$\Gamma$ transition) with $E_g$ = 0.823 eV, similar to the results of Sun et al. [39]. However, the calculated partial densities of states also demonstrated that the top of the valence band of the CIPS is created by electrons belonging to the Cu and S atoms, and the bottom of the conduction band is formed by mixed orbitals of In and $P_2S_6$ complexes in both phases. Charge distribution in para- and ferri-electric phases of the CIPS revealed electronic atom configurations: for para-phase CIPS, Cu: $3d^{9.89}4s^{0.52}4p^{0.40}$, In:$4d^{10}5s^{0.98}5p^{1.22}$, P: $3s^{1.49}3p^{2.96}$, S: $3s^{1.88}3p^{4.46}$, and for ferroelectric phase CIPS, Cu: $3d^{9.86}4s^{0.52}4p^{0.40}$, In:$4d^{10}5s^{0.90}5p^{1.20}$, $P_1$: $3s^{1.52}3p^{2.94}$, $P_2$: $3s^{1.48}3p^{2.93}$, $S_1$: $3s^{1.86}3p^{4.50}$, $S_2$: $3s^{1.89}3p^{4.50}$. And there is a tendency to increase the states mixing during the transition to ferrielectric phase. Recently, Sun and coauthors analyzed the structural transition of monolayer CIPS from the initial antiferroelectric (AFE) (**Fig. 18c**) to out-of-plane ferroelectric (FE) state under compressive strain and electric field [39], and the AFE ground state is consistent with the previous report [46]. When the strain is increased to ~ -0.6%, the FE state changes to the ground state. Compared electronic structure with that of $CuInP_2Se_6$ and $AgInP_2Se_6$ [94,95], the instability of $CuInP_2S(Se)_6$ crystals was explained by the second-order Jahn-Teller effect [46]. Based on the calculated crystalline and electronic structure, Zhang et al. studied the mechanical properties of $ABP_2X_6$ monolayer and predicted an out-of-plane negative Poisson's ratio of -0.060 [96].

### 4.2 Optical properties

The optical properties of CIPS have been widely studied using Raman spectroscopy, second harmonic generation (SHG), and UV-visible absorption spectroscopy, which revealed more information about the crystal structure and electronic properties [56,97-100]. In 1998, Vysochanskii et al. firstly measured the temperature-dependent Raman spectra of CIPS crystal and determined the order-disorder transition $T_c$ ~ 315 K [97]. It was suggested that the copper hopping motions induced ferroelectric transition, and the ionic conductivity in this material is derived from the coupling between $P_2S_6$ deformation modes and $Cu^I$ vibrations. Moreover, the temperature evolution of the phonon spectra across the phase transition exhibited relaxation phenomenon, instead of resonant behavior, which was related to the dynamic disorder of $Cu^I$ in the paraelectric phase, in agreement with other reports [32,41,62]. SHG is a versatile probe of crystallographic symmetry, especially for polar materials, which was first performed on CIPS by Misuryaev et al. in 2000 [98]. The SHG of CIPS single crystal was measured between 250 K and 335 K, revealing the phase transition temperature of ~315±2 K, consistent with other reports. In 2008, Guranich and coworkers investigated the dielectric responses and birefringence of CIPS crystal under different hydrostatic pressures [56]. The same conclusion was drawn from both studies that increasing pressure contributed to a linear increase of $T_C$, which was discussed in the previous session.

Studenyak and coauthors performed a detailed study of the UV-visible absorption spectra of CIPS in 2003 (**Fig. 19**) [99]. In the temperature range of 77 K- 250 K, the linear absorption edge was clear, and the absorption coefficient was well described by the coordinates $[\alpha(h\nu) h\nu]^2 = f(h\nu)$ (**Fig. 19a**), relating to direct allowed interband transition. But, when $T \geq 250$ K, the absorption edge started to be distinct, which was explained by the intense hopping of copper ions between multipotential wells and that penetrated into the interlayer space. In the temperature range of 325 K-573 K, the absorption edge has an exponential shape. Such absorption tails were well described by the empirical Urbach rule

$$\alpha(h\nu) = \alpha_0 exp\left[\frac{h\nu - E_0}{\omega}\right],$$

where $\omega = kT/\sigma$ is the absorption edge energy width, and $\alpha_0$, $E_0$, and $\sigma$ are empirical parameters determined from the experimental data [101]. The temperature-dependent $\omega$ (inset of **Fig. 19b**) usually reflects localized defects states induced by intrinsic point defects in ion crystals, ion migration, and ionic thermal vibration, and the degree of disordered state can be expressed by the magnitude of the localized band gap value [102]. Furthermore, as the temperature increased from $T_c$ to 573 K, the slope of the absorption coefficient kept decreasing (**Fig. 19b**), which also represented the enhancement of ionic migrations. Temperature variation of $E_g$ was analyzed based on the optical absorption behavior (**Fig. 19c**), demonstrating a typical order-disorder transition at 315 K. Such temperature-dependent $E_g$ was considered to be contributed by two aspects: lattice thermal expansion and exciton (electron)-phonon interaction (EPI). In the ferroelectric phase, the EPI dominates, while both contribute near the transition point and in the paraelectric phase.

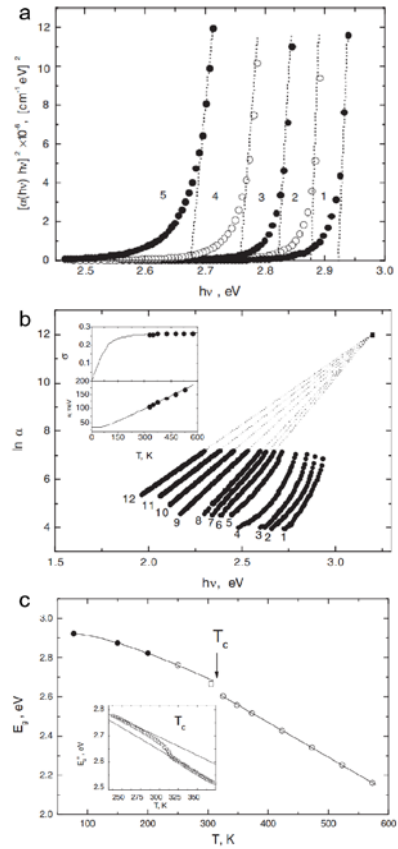

**Fig. 19 a-b,** Spectral dependencies of CPS crystal absorption coefficient at different temperatures:1, 77 K; 2, 150 K; 3, 200 K; 4, 250 K; 5, 305 K; 6, 325 K; 7, 348 K; 8, 373 K; 9, 423 K; 10, 473 K; 11, 523 K; 12, 573 K. The inset of **b** shows the temperature dependence of the exponential absorption edge energy width $w$ and $\sigma = kT/w$ parameters: the experimental values for $T>T_c$ are shown by filled circles, and those calculated by solid curves. **c,** Temperature dependence of the CIPS crystal energy gap, the experimental values of $E_g$ are shown by circles while the calculated values are represented by solid lines [99].

## 5. Devices and applications

### 5.1 Ultrasound transducer

In terms of the electromechanical coupling effect, Banys and coworkers performed a series of studies on the ultrasonic properties of CIPS crystal [64,103,104]. They observed that along the polar axis of CIPS, the longitudinal ultrasonic velocity plunges while the attenuation peaks at the phase transition. The temperature and frequency dependence of the ultrasonic behavior are consistent with those of the dielectric spectra and can be described by a relaxational soft-mode linked to the order-disorder transition. The extremely large longitudinal elastic nonlinearity in CIPS was further verified by several later studies [105-107]. It was also reported that the change of ultrasonic velocity ($\Delta V$) had a quadratic dependence on the DC bias field, and exhibited anomalies when domain switching took place. Under external DC bias along the polar axis of $CuInP_2(S,Se)_6$ thin plates, the square of electromechanical coupling coefficient ($K_{33}^2$) could be as high as 20%~30% to generate longitudinal vibrations in the paraelectric phase. The value is much higher than that in the megahertz range (~8%) [103], suggesting these materials offer possible applications for low-frequency transducers.

### 5.2 Ferroelectric capacitor and ferroelectric field effect transistor

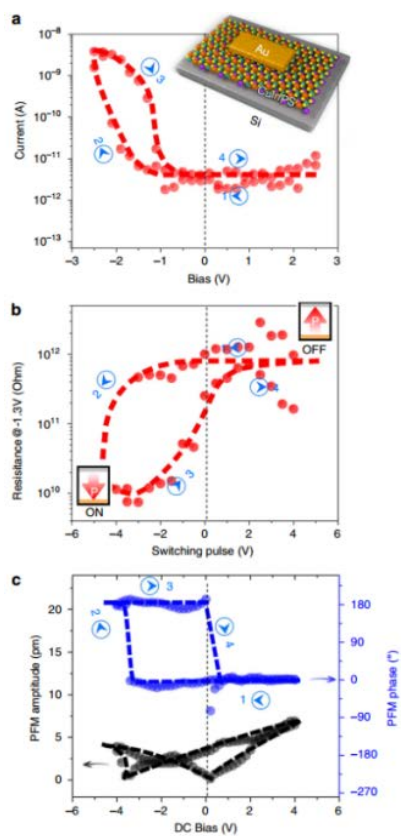

**Fig. 20 a,** The *I-V* curve of a typical vdW Au/CIPS/Si capacitor with 30 nm thick CIPS. Inset is the schematic of the device. **b,** Resistance-switching hysteresis loop of the device measured at a bias voltage of 1.3 V. **c,** Out-of-plane PFM amplitude (black) and phase (blue) loops of the same device. [13]

The most widely-pursued application of ferroelectrics is for non-volatile memories based on their switchable polarization. It manifests in a myriad of prototype devices, including ferroelectric resistive switching capacitors, ferroelectric tunnel junctions (FTJs), and ferroelectric field-effect transistors (Fe-FETs). In 2016, Liu and coworkers reported a ferroelectric capacitor consisting of a 30 nm CIPS flake on conducting Si substrate and Au top electrode (inset of **Fig. 20a**) [13]. Atomic force microscope (AFM) with a conductive tip was used to access each capacitor and clear piezoelectric hysteresis and resistive switching behaviors were observed. The coercive voltages in the PFM loops coincide with the points at which resistive switching takes place, suggesting that the ferroelectric polarization reversal in CIPS is the origin of the resistive switching of the device. The resulted on/off ratio is ~ $10^2$, which is comparable to that observed in tunnel junctions based on conventional ferroelectric oxides [108].

In recent years, vdW heterostructures consisting of 2D layered materials have attracted much attention because of the easy integration of different components without the constraint of lattice matching [109-111]. The incorporation of 2D ferroelectric CIPS in such heterostructures has led to Fe-FETs [30,112] and negative capacitance field-effect transistor (NC-FET) [10].

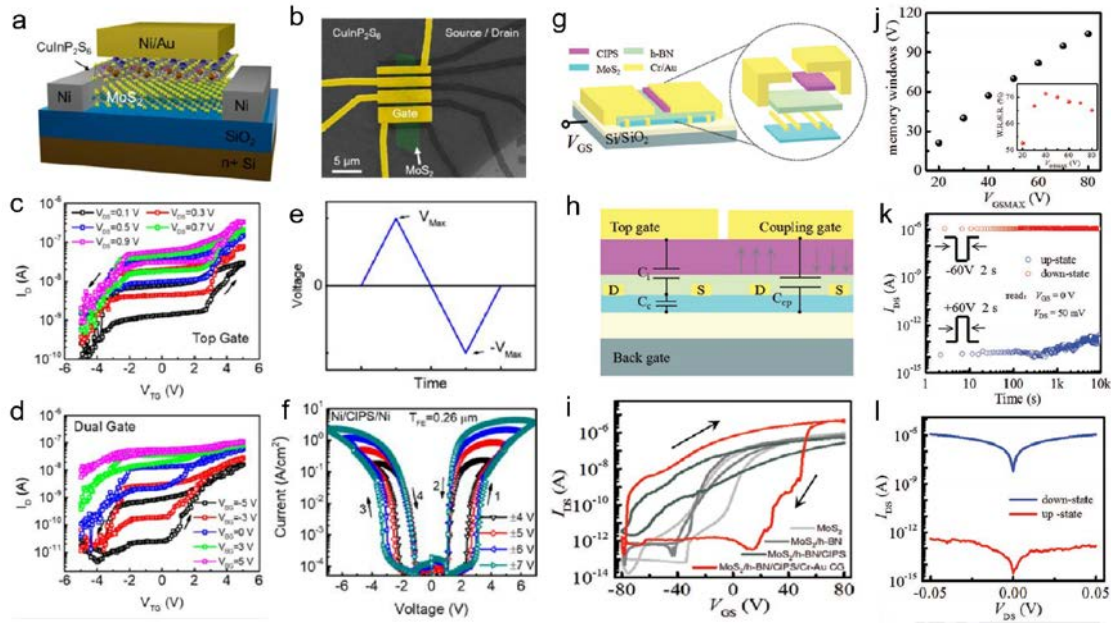

**Fig. 21 a,** Structure, and **b,** optical image of an MoS$_2$/CIPS Fe-FET. **c,** Top-gate $I_D$-$V_{GS}$ characteristic of the transistor at room temperature. **d,** Top-gate $I_D$-$V_{GS}$ characteristics at $V_{DS}$ = 0.1 V and at various $V_{BG}$ at room temperature. **e,** Illustration of the waveform used to measure the CIPS capacitor, and **f,** its I-V characteristics at different sweep ranges [30]. **g,** Structure, and **h,** equivalent circuit of the MoS$_2$/CIPS/BN device. **i,** $I_{DS}$-$V_{GS}$ characteristic, **j,** memory windows, **k,** Retention behavior, and **l,** $I_{DS}$-$V_{DS}$ characteristics under different polarization states of the device [112].

In 2008, Si et al. reported Fe-FETs based on CIPS and 2D semiconductor MoS$_2$ [30]. A 0.4 μm CIPS flake was used as the ferroelectric top gate (**Fig. 21a,b**). The channel was a 7 nm MoS$_2$ (1 μm length x 2.1 μm width). The top gate transfer characteristics at room temperature showed clear counterclockwise hysteresis loops (**Fig. 21c**), which can be modulated by the back gate bias, $V_{BG}$ (**Fig. 21d**). The authors also studied the resistive switching behavior of a metal-insulator-metal (MIM) capacitor composed of 0.26 μm thick CIPS flake and nickel as the metal electrodes (**Fig. 21e,f**). An on/off ratio of up to $10^4$ was observed.

Recently, Huang and coworkers modified the device structure by inserting a layer of hexagonal boron nitride (h-BN) in between CIPS and MoS$_2$, and the whole structure consisted of two devices (one with and the other without top CIPS layer) (**Fig. 21g**) [112]. Comparing the $I_D$-$V_{GS}$ characteristics of different setups, the MoS$_2$/h-BN/CIPS/Cr-Au top gate configuration showed the largest hysteresis loop (red line in **Fig. 21i**) and an on/off ratio as high as $10^7$. The $I_D$-$V_{GS}$ characteristic also revealed a low off-state channel current of $10^{-13}$ A when the polarization of CIPS layer is pointing up (**Fig. 21i**). Notably, this off state channel current is much smaller than some other vdW materials-based devices [113,114]. The memory window was enhanced with increasing $V_{GS}$, and the largest value was obtained when $V_{GSMAX}$ = 80 V (**Fig. 21j**). For nonvolatile memory application, the retention and cyclic endurance properties were also tested. $V_{GS}$ pulses of ± 60 V (2 s) were used to control the state of the device. While the channel current of the polarization down state remained stable for $10^4$ s, the increase of current in the polarization up state reduced the on/off ratio from $10^8$ to $10^7$ (**Fig. 21k**). These results demonstrate the potential of ferroelectric CIPS in non-volatile memories.

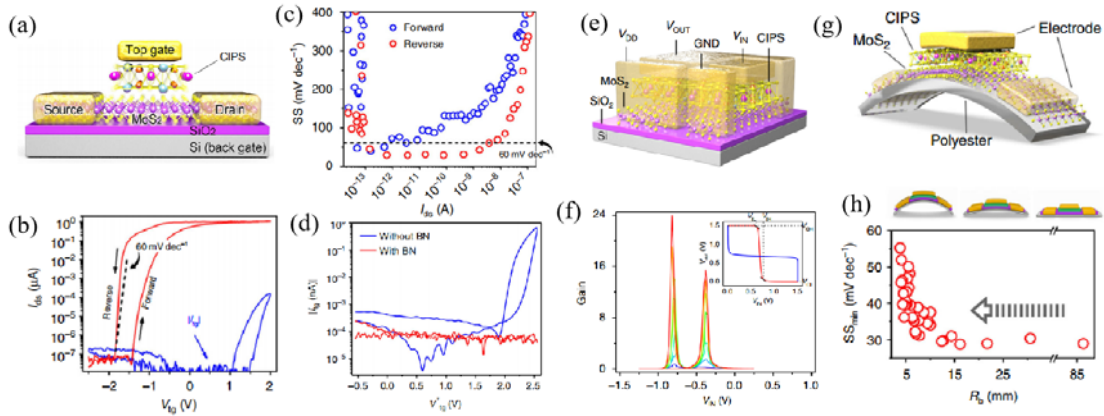

**Fig. 22 a,** Schematic of a CIPS/MoS$_2$ negative capacitance field-effect transistor (NC-FET). **b,** Top-gate $I_{ds}$-$V_{tg}$ characteristics (red) and leakage current (blue), and **c,** SS-$I_{ds}$ characteristics, of the NC-FET. **d**, Leakage currents vs. top gate bias of the NC-FETs with and without h-BN layer. **e,** Schematic illustration of a vdW NC-FET inverter. **f,** Voltage gain of the inverter at various $V_{DD}$ from 0.1 V (blue) to 1.5 V (red). **g,** Schematic illustration of a vdW NC-FET on a flexible substrate and, **h,** the effect of bending radius on SS. [10]

Negative capacitance (NC) field-effect transistor (NC-FET) has been experimentally demonstrated, which can break the thermionic limit and achieve ultra-low power consumption [115-116]. Boltzmann's limit defines that the minimum gate voltage required for increasing the $I_{ds}$ by 10 times at 300 K is 60 mV, namely 60 mV dec$^{-1}$ subthreshold swing (SS). In an NC-FET, the ferroelectric material effectively amplifies the applied gate voltage, and vdW ferroelectrics are expected to offer better performance and reliability by minimizing the dangling bonds and chemical traps at the interface [117,118]. In a recent report, Wang et al. demonstrated NC-FETs using CIPS as the ferroelectric dielectric layer and atomically thin $MoS_2$ as the semiconductor channel [10]. The top-gate $I_{ds}$-$V_{tg}$ measurement was conducted on a four-layer $MoS_2$ device with 51 nm CIPS (**Fig.22a,b**). A counterclockwise hysteresis loop with an on/off ration of $10^7$ was observed, which originated from the ferroelectric nature of CIPS. From the forward and reverse sweeps of the hysteresis, the extracted minimum SS values are 39 and 28 mV $dec^{-1}$, respectively, much lower than the thermionic limit (**Fig.22c**). The sub-60 mV $dec^{-1}$ SS was repeatedly verified in 21 devices with various CIPS thickness, revealing excellent reliability. The authors tried inserting a 7.5 nm h-BN layer into the top-gate stack, with which the $I_{ds}$-$V_{tg}$ hysteresis was effectively suppressed. The gate leakage current was also reduced by more than 3 orders of magnitude (**Fig. 22d**). A logic inverter was constructed with two CIPS/$MoS_2$ NC-FETs serving as the pull-up load and pull-down driver (**Fig. 22e**). During signal inversion, maximum voltage gain as high as 24 was obtained for the supply voltage of $V_{DD}$ = 1.5 V (**Fig. 22f**), which is higher than that of conventional inverters based on transition metal dichalcogenides (TMDs) [119]. Due to the vdW nature of the components. $MoS_2$/CIPS NC-FETs can be fabricated on a flexible substrate (**Fig.22e,f**). The transfer characteristics still exhibited counterclockwise hysteresis loop and sub-60 mV $dec^{-1}$ switching, even when the device was bent to various degrees (**Fig. 22 g,h**).

Very recently, FTJ devices based on CIPS/graphene vdW heterostructures show a stunning tunneling electroresistance (TER) above $10^7$, demonstrating the great potential of ferroelectric vdW heterostructure in resistive memories [120].

**5.3 Pyroelectric nanogenerator and electrocaloric cooling device**

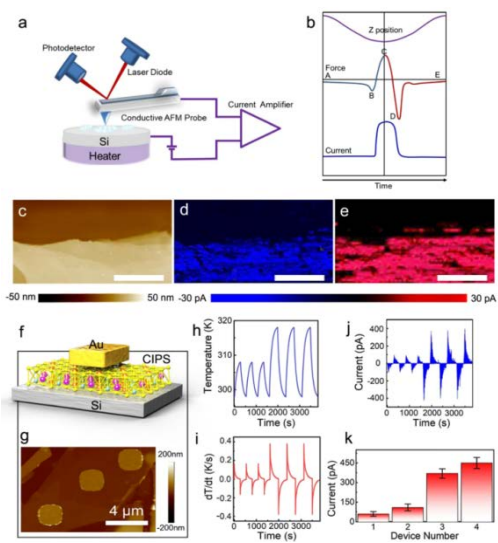

**Fig. 23 a,** Schematic illustration of Peak Force TUNA setup for simultaneous topography and electrical property mapping. **b,** Plots of Z position, force, and current as functions of time during one Peak Force Tapping cycle. **c,** AFM topography, and TUNA current map during **d,** heating from 298 K to 318 K and, **e,** cooling from 318 K to 298 K. **f-g,** Schematic and AFM image of the vdW CIPS/Si device. **h,** Temperature cycling for the pyroelectric measurement, and **i,** the corresponding differential curve. **j,** The current output of a vdW CIPS/Si device when temperature changes. **k,** The current outputs of devices with different electrode sizes. [35]

Ferroelectric materials are naturally pyroelectric and current can be generated by the change of spontaneous polarization upon heating or cooling. Since the $T_C$ of CIPS is very close to room temperature, the spontaneous polarization is sensitive to small temperature change, making it highly promising for thermal sensing applications. Recently, Niu and coworkers investigated the pyroelectric performance of CIPS based 2D devices (**Fig. 23**) [35]. CIPS nanoflakes of different thicknesses were exfoliated and placed on heavily doped Si substrates, and the temperature-dependent surface potential changes ($\Delta V$) were detected by using Kelvin probe force microscopy (KPFM). Their study demonstrates that CIPS remains pyroelectric down to a few nanometers, even for bilayer, which breaks the reported critical limit for ferroelectricity [13,34,69]. Pyroelectric currents of CIPS nanoflakes under varying temperature were measured using Bruker Peakforce TUNA modulus with a conductive probe (**Fig. 23a-e**). With Au top electrodes, it was shown that cyclic modulation of the temperature between 298 K and 308 K caused switching of output current between positive and negative signs, and the magnitude of current depended on both electrode size and the temperature ramping rate (**Fig. 23h-k**) [35].

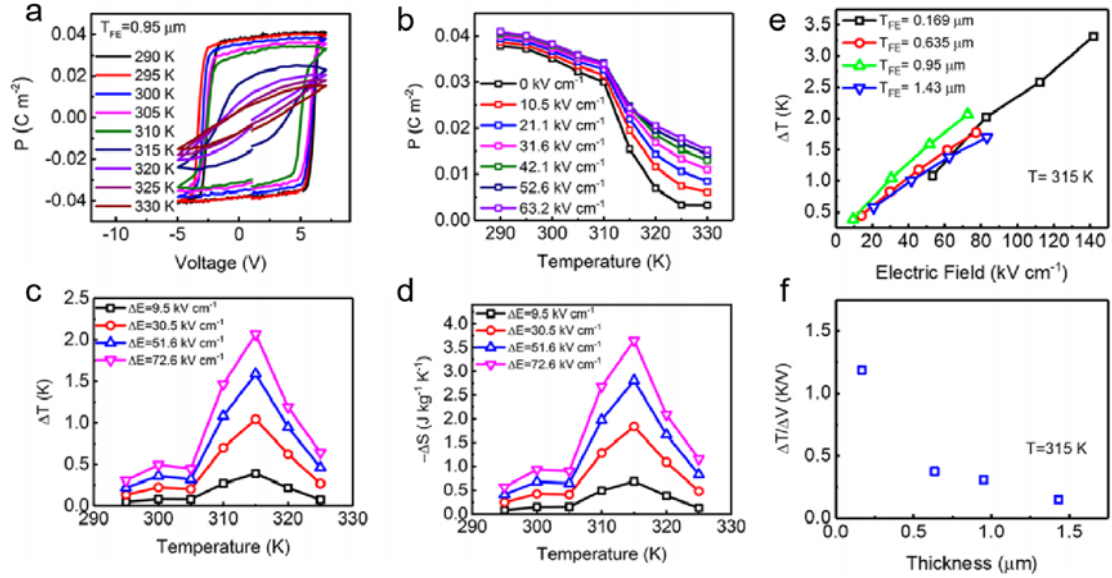

**Fig. 24 a,** *P-E* loops of a CIPS capacitor measured at different temperatures. **b,** Polarization versus temperature at different electric fields, extracted from (a). **c,** Adiabatic temperature change $|\Delta T|$ versus temperature, and **d,** isothermal entropy change $|\Delta S|$ versus temperature for different $|\Delta E|$ values. **e,** Electrocaloric strength $|\Delta T|/|\Delta E|$ of CIPS capacitors with various CIPS thicknesses. **f,** $|\Delta T|/|\Delta V|$ of different CIPS capacitors. [36]

Electrocaloric (EC) effect refers to the change in temperature of a material upon application or withdrawal of an electric field under adiabatic conditions. The two representative characteristics, adiabatic temperature change $|\Delta T|$ and isothermal entropy change $|\Delta S|$, can be evaluated using the following equations [121]:

$$|\Delta T| = -\int_{E_1}^{E_2} \frac{1}{C_p} T \left(\frac{\partial P}{\partial T}\right)_E dE \quad \text{and} \quad |\Delta S| = \int_{E_1}^{E_2} \frac{1}{\rho} \left(\frac{\partial P}{\partial T}\right)_E dE, \text{ where } C \text{ is the heat capacity and } \rho$$

is the density of the EC material. Si and coworkers investigated the EC effect of CIPS in 2019 [36]. Polarization-electric field loops at different temperatures were measured, from which polarization vs temperature relationships under various electric field were extracted (**Fig. 24a,b**) and $|\Delta T|$, $|\Delta S|$ calculated (**Fig. 24c,d**). For a capacitor with 0.95 μm thick CIPS, the calculated maximum |Δ$T$| was 2.0 K, and the maximum |Δ$S$| is 3.5 J kg$^{-1}$ K$^{-1}$ at 315 K for |Δ$E$| =72.6 kV cm$^{−1}$. As the thickness of CIPS flake decreases, the |Δ$T$|/|Δ$V$| value increases fast, while the relative EC strength (|Δ$T$|/|Δ$E$|) remains relatively stable (**Fig. 24e,f**) [36]. Therefore, 2D ferroelectrics such as CIPS may find potential applications in solid-state refrigeration, especially for micro- and/or nano-scale devices [122,123].

Recently, Duan and coauthors theoretically constructed a CIPS/MnPS$_3$ heterostructure and studied the valley polarization state of CIPS in such a ferroelectric/antimagnetic system [124]. By first-principles calculations, a valley splitting was predicted to occur in CIPS because of the magnetic proximity effect induced time-inversion symmetry breaking. What's more, the valley degree of freedom in this system can be controlled by the polarization direction of CIPS, which may be a promising candidate for all-electrically reading and writing memory devices.

## 6. Open questions

The renewed interest in CIPS in the past few years has led to new discoveries in both fundamental physics and applications. Learning from the success of BiFeO$_3$ in the field of multiferroic research, one can envision CIPS as the archetype in its homologous TPS family, which hosts a broad spectrum of versatile properties and potential functionalities. Furthermore, the capability of downscaling such 2D materials holds great promises for device miniaturization and flexibility. Herein, we would like to point out some open questions and intriguing directions regarding CIPS and related materials.

**Multistate ferroelectricity**. The interesting properties of CIPS largely stem from the coordination instability of Cu, which results in multiple metastable sites in the crystal lattice. Both theoretical calculations and experiments have identified an additional metastable site within the vdW gap, which is characterized by a greatly enhanced polarization (~12 μC/cm$^2$) compared to that of the ground state (~4 μC/cm$^2$). However, the potential wells are shallow with a small energy difference on the order of tens of meV, comparable to the thermal energy at room temperature [40]. Therefore, the occupancies of the Cu cations over the different sites would appear as smeared lobes instead of discrete states, as evidenced in the electron density maps. Recently, Brehm et al. from ORNL reported experimental observation and controlled switching over the quadrupole polarization states using PFM [31]. It should be pointed out that mixed-phase CIPS-IPS sample was used in this study. Hence, it is possible that the metastable in-gap state was stabilized by the heterostructure strain and domain walls, leading to multiple switchable polarization states. Certainly, further studies are required to clarify this issue. Given the ionic conductivity of CIPS, the deterministic control of polarization switching between multiple states, while avoiding disordered ionic migration would then be crucial to the multi-level memory applications. The intricacy between polarization switching and ionic migration may bring about other unprecedented behaviors in CIPS, such as the out-of-plane polarization switching by in-plane field [67] and alignment of polarization against an electric field [125].

**Piezoelectric and electrochemical strain**. Although the negative longitudinal piezoelectric effect of CIPS has been quantitatively studied, its tensorial piezoelectric responses remain unexplored.

This is closely related to its mechanical parameters. The transverse piezoelectric coefficients deduced from the spontaneous lattice strain across the $T_C$ are all negative (**Fig. 16c**), which suggests CIPS could be an unusual case of electric-auxetic materials with negative Poisson's ratio. Furthermore, local PFM study has shown considerable electrochemical strain above $T_C$ (**Fig. 8**), which is probably inevitable at room temperature as well. To deconvolute different electromechanical coupling mechanisms pertinent to the ionic dynamics, ultrafast synchrotron diffraction experiments with in-situ electric field and microscopy capability are highly desirable.

**Hetero-epitaxial CIPS-IPS composite.** The discovery of spontaneous phase separation in Cu-deficient CIPS provides a new playground for tuning its properties. Parameters, such as composition, hetero-interfacial strain, domain size, can then be engineered to tailor the ferroelectric, mechanical, optical properties of the composite crystals. The complex hierarchical domain structure and the abundance of the domain boundaries may host emergent functionalities and potential topological structures. Again, these naturally-formed heterostructures can exist at nanoscale thickness, which is ideal for 2D electronics and optoelectronics devices.

**Twisting the 2D ferroelectricity.** Thanks to the weak interlayer interaction, one crystal layer in vdW materials can be rotated with respect to the adjoining layer to form Moiré superlattice with various periodicity depending on the twisting angle. This consequently modifies the interlayer bonding and electronic structures, resulting in emergent phenomena. Inspired by the rich physics uncovered in twisted graphene and other 2D materials, the question naturally arises as how this effect may influence the ferroelectricity in 2D materials. In CIPS, the second-order Jahn-Teller effect of the Cu-S sublattice signifies that the interlayer Cu-S interaction plays a role in stabilizing the ferroelectric order. A twist in the adjacent layers would possibly disrupt the Jahn-Teller ground state. Whether it will enhance or decrease the potential double well, or even produce multiple wells, would be an appealing topic to investigate in the future. Besides twisting, stacking pattern is also crucial to the properties of 2D materials. A recent study shows a rearrangement in the stacking pattern of CIPS below a critical thickness [23]. While the result requires further verification, it certainly implies other thermodynamically stable polymorphs, in analogous to the 2H and 3R phases of metal dichalcogenides.

There are so many interesting topics in the field of CIPS and the TPS family, though we are only able to name a few here. Other open questions include: high-pressure phase transition in CIPS, the exploration for multiferroism in TPS compounds with strong magnetoelectric coupling, excitonic physics and its interaction with ferroic orders, etc. Hopefully, this review can spark the imaginations of researchers in the relevant communities to unveil new physics and technological applications of CIPS and related materials.

**Acknowledgement**

This review is supported by Natural Science Foundation (NSF) of China (11904176, 61874060, 61911530220, U1932159,11774249), NSF of Jiangsu Higher Education Institutions

(19KJB140004), the startup found from NJUPT (NY219028, NY217118), NSF of Jiangsu Province (BK20181388, BK20171209), and the Key University Science Research Project of Jiangsu Province (18KJA140004). The authors also gratefully acknowledge the startup fund from Soochow University, Priority Academic Program Development (PAPD) of Jiangsu Higher Education Institutions, Jiangsu Specially-Appointed Professor program, and the startup grant from Southern University of Science and Technology.